\documentclass[12pt]{article}
\usepackage{a4}
\usepackage[utf8]{inputenc}
\usepackage{amsmath,amsfonts,graphicx,pdflscape,booktabs,caption,subcaption}
\usepackage{setspace}

\newcommand{\var}{\operatorname{var}}

\newtheorem{example}{Example}

\title{Supplementary Material for\\
``Should we sample a time series more frequently?\\
Decision support via multirate spectrum estimation (with discussion)''}

\author{Guy P. Nason\thanks{University of Bristol, UK}, Ben Powell\footnotemark[1], Duncan Elliott\thanks{Office for National Statistics, Newport, UK} and Paul A. Smith\thanks{University of Southampton, UK}}

\begin{spacing}{1}
\begin{document}

\maketitle

\begin{abstract}

This technical report includes an assortment of technical details and extended discussions related to paper
``Should we sample a time series more frequently? Decision support via multirate spectrum estimation
(with discussion)'', which introduces a model for estimating the log-spectral density of a stationary discrete time process given systematically missing data and models the cost implication for changing the sampling rate.
\end{abstract}

\tableofcontents

\section{Investigating the log-likelihood surface for AR parameters given systematically missing data}

\subsection{Context}
This section describes a tangential investigation from the main project, in which we sought to advise research partners whether or not to increase the sampling rate of an economic variable on the basis of a long history of observations at a low rate and a short trail series of observations at a higher rate. We then, in fact, introduced a third option: a recommendation to postpone the first decision and to collect more high frequency data. The precise specification of the decision problem and our solution to it are described in \cite{NPES15}. As a more general observation, we noticed that our decision is highly sensitive to estimates for the process's log-spectrum and quantifications for their uncertainty. Importantly, we also observed that standard estimates for these objects derived from fitting ARMA models are frequently very poor, a statement we can qualify by referring to the sub-optimal decisions they lead to. In this section, we present a brief exploration of what is going wrong with the ARMA fitting 
procedure.

\subsection{Theory}\label{supplementarytheory}
From our knowledge of aliasing, we know that there is an intrinsic ambiguity for a spectral density constrained by observations of a stationary discrete-time process taken at every other time point, which we will call $D_{low}$. The ambiguity comes in the form of a symmetry about $\omega_{Nyquist}=1/4$. We can infer, for example, that there is an accumulation of spectral power around $\omega=1/8$ or around $\omega=3/8$ but we cannot infer the ratio of power at these locations. This is problematic for a computer program that tries to fit an ARMA model to the data via optimization of a likelihood function because the parametric parsimony that characterizes these models endows them with a tendency to assume the spectral power is highly localized. An AR(p) model can encode a spectrum with no more than $\lceil p/2 \rceil$ modes, for example.

The result, in general, is that in such a situation an optimizer will rush to one conclusion or the other: that the spectral power is in the lower- or upper-half of the spectral domain. Additionally, preferences for parsimony, typically quantified via the AIC or BIC, will tend to suppress the possibility that there are accumulations of power at both places, which is a conclusion requiring the introduction of more ARMA parameters.

The symmetry, and associated confounding, described here is broken with the introduction of a series of high frequency observations of the same process, which we will call $D_{high}$. As this second dataset gets larger, the formerly symmetric likelihood surface for the AR parameters begins to deform: the heights and shapes of the likelihood's modes are adjusted, with one eventually flattening out altogether if the spectral power is indeed located entirely at one of the aliased frequencies. An interesting question to ask is how quickly this deformation occurs and how successfully it steers optimization routines from rushing to the wrong conclusion.

\subsection{Simulation experiment}
\begin{figure}
\centering
\includegraphics[width=1\textwidth]{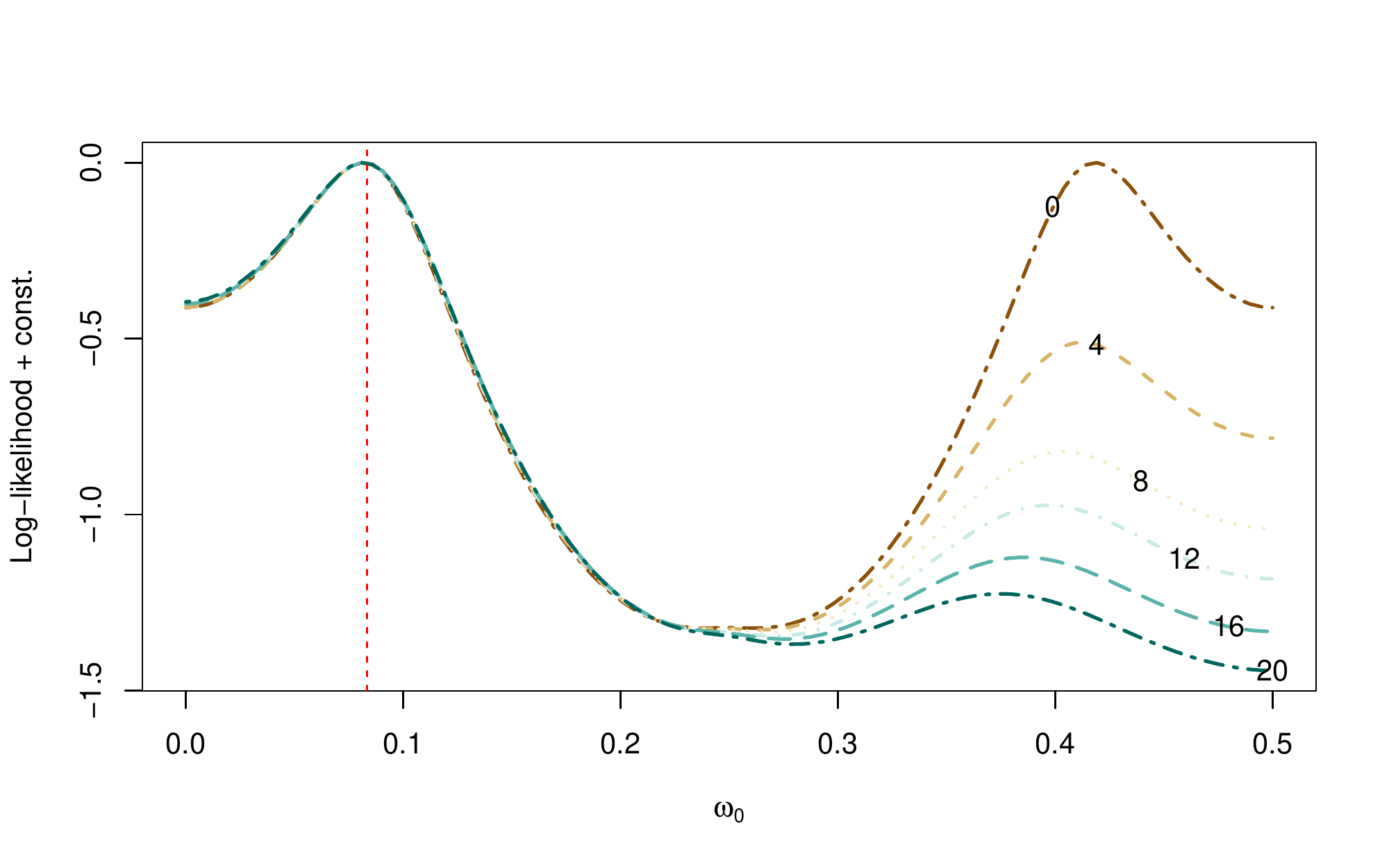}
\caption{\label{loglikplot} Each curve here describes a Monte Carlo
average of log-likelihood surfaces for the parameter $\omega_{0}$, which
defines a pair of AR coefficients according to \eqref{zz}. The red vertical dashed line marks the location of $\omega_{0}$'s true value. The surfaces, which we have labeled, correspond
to ensembles of simulations for which the value of $N_{low}=128$ and $N_{high}=0,4,8,\ldots,20$. Note also that each surface has had its maximum subtracted in order to align them.}
\end{figure}

Figure \ref{loglikplot} illustrates the outcome of an numerical experiment in which we fit an AR model to $N_{low}=128$ observations at every other time point and $N_{high}$ subsequent observations at every time point. These observations are simulated using the AR(2) model
\begin{align*}
 x_{t}=\phi_{1}x_{t-1}+\phi_{2}x_{t-2}+e_{t},
\end{align*}
where $e_{t} \sim \mathcal{N}(0,\sigma_{innov.}^2)$ and the model parameters are given by
\begin{align}\label{zz}
\phi_{1}=z+\bar{z}, && \phi_{2}=-z\bar{z}, && \sigma_{innov.}^2=1,\\
z=0.9 \exp(-\imath 2 \pi \omega_{0}).
\end{align}
Parameterization in terms of the complex number $z$ allows us to focus on its argument, $\omega_{0}$, which corresponds to a peak in the process's spectral density. Fixing the modulus of $z$ at $0.9$ makes for a distinct peak and allows us to plot one dimensional log-likelihood functions which are much easier to comprehend and to communicate.

After simulating the time series, $D_{high}$ and $D_{low}$, from the AR model with $\omega_{0}=1/12$, we compute the log-likelihood for the parameter $\omega_{0}$ at a set of equally spaced values. This procedure is iterated over $500$ simulated data sets to produce an average vector of log-likelihoods. We then repeat the whole procedure for different values for $N_{high}$.

In Figure \ref{loglikplot} we show the evolution of the average log-likelihood surface as more high-frequency data is received. As expected, more high frequency data serves to diminish the sub-optimal mode, and we observe that at least 20 high-frequency observations are required before we can be reasonably confident there is not a danger of an optimizer being tricked into settling there.

\begin{figure}
\centering
\includegraphics[width=1\textwidth]{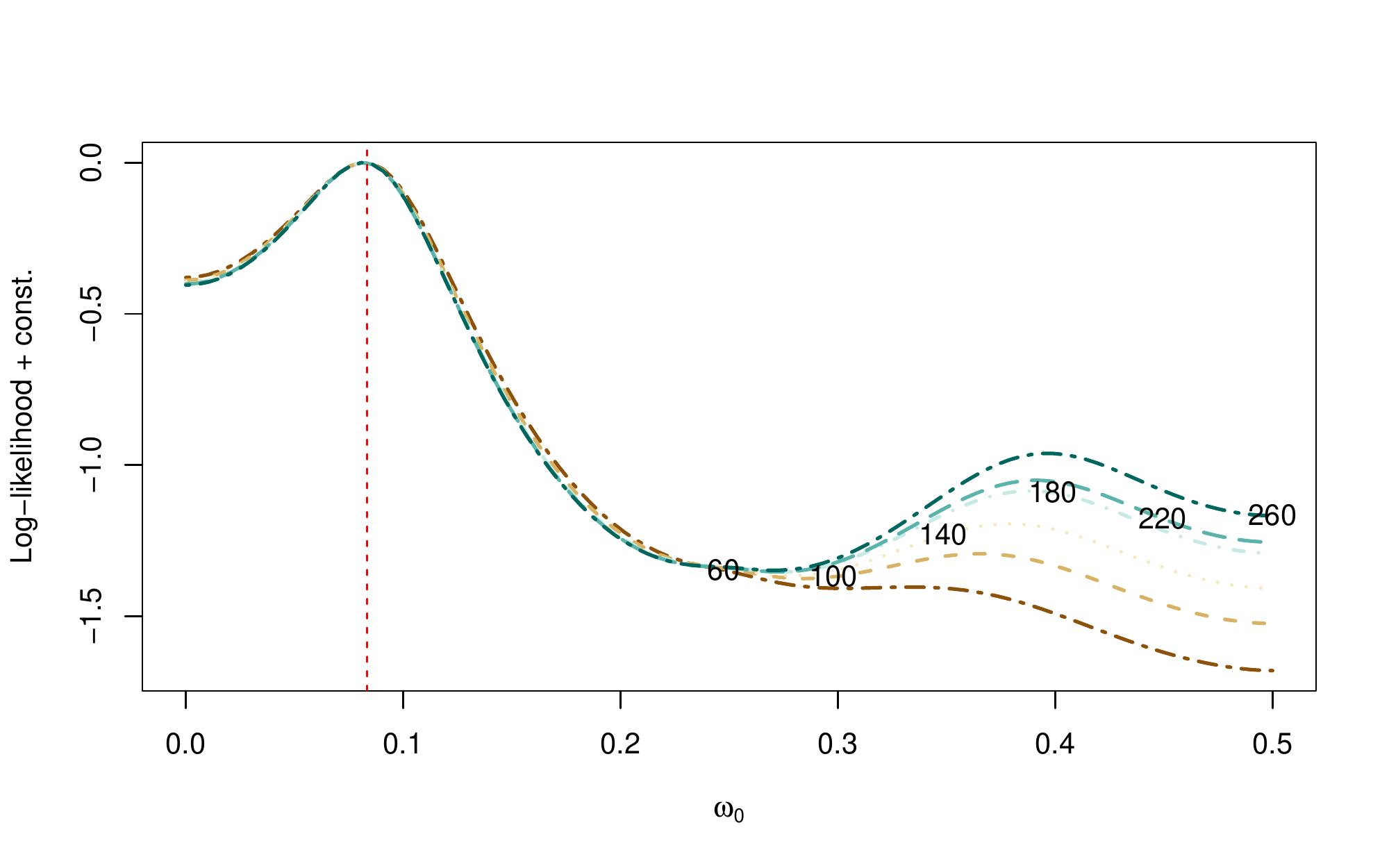}
\caption{\label{loglikplot2} As in Figure \ref{loglikplot}, the curves here describe Monte Carlo
average of log-likelihood surfaces for the parameter $\omega_{0}$ aligned by their maxima. Now the lines correspond
to ensembles of simulations for which the value of $N_{high}=20$ and $N_{low}=60,100,140,\ldots,260$.}
\end{figure}

In a second simulation experiment we perform the same calculations but with $N_{high}$ fixed at $20$ and $N_{low}$ set to one of a sequence of values. Interestingly, as illustrated in Figure \ref{loglikplot2}, we notice that both modes become more pronounced when $N_{low}$ is increased, and that as a consequence having more data can in fact make the maximum likelihood procedure more susceptible to finding the wrong mode and producing misleading inferences.

\subsection{Discussion}
For the cases in which we identify two modes, it is not accurate to conclude that the maximum likelihood calculation leads to misleading inferences half of the time. The optimizer's steps, and most likely our specification of the optimizer's initial conditions, are deterministic, meaning that the calculation produces those misleading inferences every time or none of the time. Accordingly, we cannot meaningfully report an expected probability of error, nor an expected size of error, for the spectrum estimated via likelihood optimization.

While this example is in one sense limited in scope, since we look at one model and one aliasing scenario, its lessons ought to be clear. The multi-modality of the likelihood induced by subsampled data opens a distinct vulnerability for maximum likelihood fitting procedures. The problem can only get worse when the thinning to produce $D_{low}$ is more severe since there are more symmetries to contend with. For instance, an optimizer trying to fit a spectrum with $m$ peaks, using a data set of observations at every $K^{th}$ time point, will have $K^{m}$ likelihood modes to choose from, corresponding to the $K^{m}$ possible combinations of peak locations. The situation becomes even worse when we decide to account for uncertainty for the number of modes, or the ARMA model's order. If we entertain the idea that an accumulation of spectral power might be distributed evenly between a subset of aliased frequencies, rather than being concentrated at just one, we now have $(2^{K}-1)^{p}$ peak combinations and 
likelihood modes to hunt down. In practice AIC-type criteria guide users away from concluding that the spectral power is divided between aliased frequencies, thus steering them away from spectra/processes that are more complex in terms of their ARMA parameterization. But when the ARMA framework is acknowledged as a practical computational tool rather than a genuine expression of belief relating to causal mechanisms contributing to the observed process, one must question the AIC's authority.

A pragmatic response to the problem of multi-modality is to re-initialize the optimizer at a set of well-spaced points in the ARMA parameter space. While we acknowledge that this is a sensible option, it does significantly increase the computational demand of the fitting procedure. Moreover, this way of searching for modes ignores the fact that we can anticipate their locations due to our understanding of the aliasing phenomenon that causes them. Indeed, the aliasing theory reveals to us an important aspect of the likelihood's global structure, without which local optimization routines must search blindly.

\subsection{A solution}
The linear Bayes estimate of the log-spectrum described in \cite{NPES15} overcomes the problems of likelihood multi-modality and of ARMA order selection. Key to its ability to deal with the first problem, is the explicit recognition of the aliasing phenomenon in its parameter adjustment equations. Issues regarding ARMA order are diminished since our method places a soft constraint on the spectrum where the ARMA model places hard constraints; by which we mean, for example, that an ARMA model of given order cannot describe a spectrum with more than a small number of modes/features but the non-parametric nature of our model means that, while resisting, it can.

\section{Further examples of log-spectrum inference}
Here, we provide further examples of log-spectrum inference using simulated data. Specifically, we investigate our method's performance in response to different types of data generating process, and given data sets of varying size and missingness pattern. As a final exercise in this section, we look at the relative performance of a spectral estimation procedure based on an initial interpolation of available data.

\subsection{Testing model performance given different data abundances}
In this subsection, we look more closely at an aspect of our model-derived estimator for the log-spectrum of a process as more data is observed. We do so by generating a large set of log-spectra, each one encoding a process with different roughness and periodicity properties. By generating multivariate normal deviates with variance matrices derived from the the log-spectra, we simulate a corresponding set of time series, subsets of which are used to inform a range of Bayes linear adjusted expectations for the log-spectra (as described at length in \cite{NPES15}). We then score the estimates according to the function
\begin{align}
d_m=N_{\omega}^{-1} \sum_{i=1}^{N_{\omega}} \{ \log[f(\omega_{i})]-\log[\hat{f}_m (\omega_{i})] \}^2,\\ \omega_{j}=0,\frac{1}{2(N_{\omega}-1)},\frac{2}{2(N_{\omega}-1)},\ldots,\frac{1}{2}, \label{metric}
\end{align}
with $N_{\omega}=128$. Table \ref{bigtab} contains the mean values of the estimators' scores over a set of 1000 randomly generated processes for different combinations of series length and sub-sampling frequency.

\begin{landscape}
{%
\newcommand{\mc}[3]{\multicolumn{#1}{#2}{#3}}
\begin{table}
\centering
\resizebox{\columnwidth}{!}{
\begin{tabular}{lll|llllllllllllllllllllllll|}
&&& \multicolumn{22}{l}{$D_{2}$}\\
&&& \multicolumn{4}{l}{$\delta=1$} &\multicolumn{4}{l}{$\delta=2$} & \multicolumn{4}{l}{$\delta=3$} & \multicolumn{4}{l}{$\delta=4$} & \multicolumn{4}{l}{$\delta=5$} & \multicolumn{4}{l}{$\delta=6$} \\
&&N& 16 & 32 & 64 & 128 & 16 & 32 & 64 & 128 & 16 & 32 & 64 & 128 & 16 & 32 & 64 & 128 & 16 & 32 & 64 & 128 & 16 & 32 & 64 & 128\\ 
  \hline
$D_{1}$&$\delta=1$&16 & 0.97 & 0.79 & 0.60 & \mc{1}{l|}{0.37} & 0.95 & 0.94 & 0.78 & \mc{1}{l|}{0.73} & 1.05 & 1.00 & 0.92 & \mc{1}{l|}{0.88} & 1.08 & 0.99 & 0.93 & \mc{1}{l|}{0.89} & 1.09 & 1.06 & 0.97 & \mc{1}{l|}{0.93} & 1.15 & 1.13 & 1.09 & 1.04 \\ 
&&32 & 0.72 & 0.61 & 0.51 & \mc{1}{l|}{0.34} & 0.73 & 0.71 & 0.61 & \mc{1}{l|}{0.53} & 0.76 & 0.70 & 0.65 & \mc{1}{l|}{0.61} & 0.76 & 0.73 & 0.69 & \mc{1}{l|}{0.65} & 0.78 & 0.74 & 0.69 & \mc{1}{l|}{0.66} & 0.74 & 0.74 & 0.71 & 0.67 \\ 
&&64 & 0.44 & 0.38 & 0.35 & \mc{1}{l|}{0.27} & 0.41 & 0.41 & 0.34 & \mc{1}{l|}{0.29} & 0.44 & 0.41 & 0.36 & \mc{1}{l|}{0.34} & 0.45 & 0.41 & 0.39 & \mc{1}{l|}{0.36} & 0.45 & 0.43 & 0.41 & \mc{1}{l|}{0.38} & 0.44 & 0.43 & 0.41 & 0.39 \\ 
&&128 & 0.27 & 0.24 & 0.24 & \mc{1}{l|}{0.20} & 0.27 & 0.28 & 0.24 & \mc{1}{l|}{0.21} & 0.30 & 0.29 & 0.27 &\mc{1}{l|}{0.26} & 0.29 & 0.28 & 0.27 & \mc{1}{l|}{0.26} & 0.29 & 0.29 & 0.28 & \mc{1}{l|}{0.26} & 0.29 & 0.29 & 0.28 & 0.27 \\ 
\cline{4-27}
&$\delta=2$& 16 & 1.08 & 0.85 & 0.62 & \mc{1}{l|}{0.36} & 1.41 & 1.36 & 1.22 & \mc{1}{l|}{1.19} & 1.34 & 1.25 & 1.16 & \mc{1}{l|}{1.06} & 1.46 & 1.41 & 1.37 & \mc{1}{l|}{1.32} & 1.54 & 1.50 & 1.31 & \mc{1}{l|}{1.28} & 1.53 & 1.58 & 1.54 & 1.49 \\ 
&&32 & 1.08 & 0.86 & 0.61 & \mc{1}{l|}{0.37} & 1.36 & 1.32 & 1.20 & \mc{1}{l|}{1.19} & 1.28 & 1.15 & 1.03 & \mc{1}{l|}{0.89} & 1.43 & 1.36 & 1.31 & \mc{1}{l|}{1.27} & 1.36 & 1.31 & 1.21 & \mc{1}{l|}{1.10} & 1.49 & 1.49 & 1.44 & 1.40 \\ 
&&64 & 0.97 & 0.83 & 0.59 & \mc{1}{l|}{0.35} & 1.27 & 1.27 & 1.19 & \mc{1}{l|}{1.19} & 1.12 & 1.00 & 0.90 & \mc{1}{l|}{0.78} & 1.31 & 1.28 & 1.25 & \mc{1}{l|}{1.22} & 1.18 & 1.14 & 1.06 & \mc{1}{l|}{0.96} & 1.32 & 1.34 & 1.32 & 1.29 \\ 
&&128 & 0.92 & 0.81 & 0.57 & \mc{1}{l|}{0.37} & 1.21 & 1.21 & 1.17 & \mc{1}{l|}{1.17} & 1.02 & 0.93 & 0.84 & \mc{1}{l|}{0.72} & 1.22 & 1.21 & 1.20 & \mc{1}{l|}{1.20} & 1.08 & 1.04 & 0.98 & \mc{1}{l|}{0.89} & 1.23 & 1.25 & 1.25 & 1.23 \\ 
\cline{4-27}
&$\delta=3$&16 & 1.20 & 0.85 & 0.65 & \mc{1}{l|}{0.37} & 1.39 & 1.27 & 1.05 & \mc{1}{l|}{0.92} & 1.63 & 1.60 & 1.49 & \mc{1}{l|}{1.41} & 1.51 & 1.36 & 1.26 & \mc{1}{l|}{1.20} & 1.71 & 1.62 & 1.50 & \mc{1}{l|}{1.44} & 1.69 & 1.69 & 1.65 & 1.57 \\ 
&&32 & 1.11 & 0.83 & 0.63 & \mc{1}{l|}{0.36} & 1.27 & 1.18 & 1.01 & \mc{1}{l|}{0.91} & 1.55 & 1.55 & 1.47 & \mc{1}{l|}{1.40} & 1.42 & 1.23 & 1.17 & \mc{1}{l|}{1.10} & 1.51 & 1.45 & 1.40 & \mc{1}{l|}{1.29} & 1.60 & 1.62 & 1.58 & 1.54 \\ 
&&64 & 1.10 & 0.85 & 0.64 & \mc{1}{l|}{0.37} & 1.22 & 1.11 & 0.92 & \mc{1}{l|}{0.83} & 1.50 & 1.51 & 1.47 & \mc{1}{l|}{1.40} & 1.36 & 1.18 & 1.11 & \mc{1}{l|}{1.04} & 1.43 & 1.36 & 1.35 & \mc{1}{l|}{1.22} & 1.53 & 1.54 & 1.52 & 1.49 \\ 
&&128 & 1.06 & 0.84 & 0.65 & \mc{1}{l|}{0.39} & 1.15 & 1.02 & 0.88 & \mc{1}{l|}{0.83} & 1.43 & 1.45 & 1.43 & \mc{1}{l|}{1.39} & 1.29 & 1.14 & 1.04 & \mc{1}{l|}{0.97} & 1.37 & 1.31 & 1.32 & \mc{1}{l|}{1.19} & 1.45 & 1.45 & 1.44 & 1.43 \\ 
\cline{4-27}
&$\delta=4$&16 & 1.23 & 0.89 & 0.67 & \mc{1}{l|}{0.38} & 1.52 & 1.47 & 1.30 & \mc{1}{l|}{1.24} & 1.58 & 1.48 & 1.32 & \mc{1}{l|}{1.16} & 1.73 & 1.65 & 1.59 & \mc{1}{l|}{1.54} & 1.74 & 1.70 & 1.58 & \mc{1}{l|}{1.45} & 1.78 & 1.78 & 1.70 & 1.64 \\ 
&&32 & 1.27 & 0.95 & 0.71 & \mc{1}{l|}{0.40} & 1.57 & 1.51 & 1.35 & \mc{1}{l|}{1.26} & 1.58 & 1.44 & 1.26 & \mc{1}{l|}{1.13} & 1.74 & 1.67 & 1.61 & \mc{1}{l|}{1.56} & 1.63 & 1.58 & 1.44 & \mc{1}{l|}{1.31} & 1.74 & 1.75 & 1.68 & 1.63 \\ 
&&64 &1.21 & 0.96 & 0.71 & \mc{1}{l|}{0.41} & 1.49 & 1.46 & 1.35 & \mc{1}{l|}{1.26} & 1.45 & 1.37 & 1.22 & \mc{1}{l|}{1.09} & 1.63 & 1.59 & 1.56 & \mc{1}{l|}{1.54} & 1.54 & 1.51 & 1.44 & \mc{1}{l|}{1.31} & 1.62 & 1.63 & 1.62 & 1.59 \\ 
&&128 & 1.20 & 0.98 & 0.75 & \mc{1}{l|}{0.45} & 1.46 & 1.45 & 1.38 & \mc{1}{l|}{1.28} & 1.39 & 1.25 & 1.12 & \mc{1}{l|}{1.04} & 1.57 & 1.58 & 1.57 & \mc{1}{l|}{1.55} & 1.40 & 1.39 & 1.34 & \mc{1}{l|}{1.21} & 1.53 & 1.53 & 1.51 & 1.49 \\ 
\cline{4-27}
&$\delta=5$&16 & 1.21 & 0.89 & 0.67 & \mc{1}{l|}{0.38} & 1.47 & 1.36 & 1.20 & \mc{1}{l|}{1.13} & 1.59 & 1.54 & 1.42 & \mc{1}{l|}{1.30} & 1.65 & 1.55 & 1.49 & \mc{1}{l|}{1.44} & 1.84 & 1.84 & 1.75 & \mc{1}{l|}{1.70} & 1.81 & 1.83 & 1.79 & 1.70 \\  
&&32 & 1.21 & 0.92 & 0.68 & \mc{1}{l|}{0.39} & 1.42 & 1.32 & 1.01 & \mc{1}{l|}{0.99} & 1.59 & 1.45 & 1.30 & \mc{1}{l|}{1.25} & 1.56 & 1.40 & 1.32 & \mc{1}{l|}{1.24} & 1.78 & 1.79 & 1.73 & \mc{1}{l|}{1.71} & 1.63 & 1.66 & 1.66 & 1.56 \\ 
&&64 & 1.15 & 0.91 & 0.68 & \mc{1}{l|}{0.40} & 1.34 & 1.26 & 0.99 & \mc{1}{l|}{0.95} & 1.50 & 1.33 & 1.21 & \mc{1}{l|}{1.15} & 1.48 & 1.38 & 1.30 & \mc{1}{l|}{1.22} & 1.68 & 1.70 & 1.68 & \mc{1}{l|}{1.68} & 1.52 & 1.56 & 1.56 & 1.48 \\ 
&&128 & 1.31 & 1.07 & 0.80 & \mc{1}{l|}{0.49} & 1.48 & 1.38 & 1.06 & \mc{1}{l|}{1.03} & 1.67 & 1.48 & 1.33 & \mc{1}{l|}{1.24} & 1.62 & 1.49 & 1.41 & \mc{1}{l|}{1.30} & 1.80 & 1.80 & 1.77 & \mc{1}{l|}{1.75} & 1.65 & 1.66 & 1.63 & 1.50 \\ 
\cline{4-27}
&$\delta=6$&16 & 1.35 & 0.93 & 0.71 & \mc{1}{l|}{0.38} & 1.69 & 1.59 & 1.36 & \mc{1}{l|}{1.29} & 1.81 & 1.73 & 1.60 & \mc{1}{l|}{1.48} & 1.88 & 1.76 & 1.66 & \mc{1}{l|}{1.59} & 2.02 & 1.96 & 1.78 & \mc{1}{l|}{1.71} & 2.00 & 2.00 & 1.90 & 1.83 \\ 
&&32 & 1.33 & 0.95 & 0.74 & \mc{1}{l|}{0.39} & 1.65 & 1.57 & 1.39 & \mc{1}{l|}{1.31} & 1.75 & 1.73 & 1.62 & \mc{1}{l|}{1.50} & 1.79 & 1.68 & 1.61 & \mc{1}{l|}{1.56} & 1.89 & 1.79 & 1.69 & \mc{1}{l|}{1.57} & 1.92 & 1.94 & 1.88 & 1.83 \\ 
&&64 & 1.29 & 0.96 & 0.76 & \mc{1}{l|}{0.40} & 1.57 & 1.52 & 1.40 & \mc{1}{l|}{1.32} & 1.67 & 1.67 & 1.60 & \mc{1}{l|}{1.50} & 1.71 & 1.62 & 1.57 & \mc{1}{l|}{1.53} & 1.75 & 1.69 & 1.67 & \mc{1}{l|}{1.57} & 1.82 & 1.85 & 1.83 & 1.81 \\ 
&&128 & 1.37 & 1.07 & 0.86 & \mc{1}{l|}{0.48} & 1.66 & 1.60 & 1.49 & \mc{1}{l|}{1.38} & 1.74 & 1.73 & 1.66 & \mc{1}{l|}{1.55} & 1.78 & 1.66 & 1.63 & \mc{1}{l|}{1.58} & 1.80 & 1.72 & 1.70 & \mc{1}{l|}{1.60} & 1.87 & 1.88 & 1.86 & 1.84 \\ 
\hline
\end{tabular}
}
\caption{\label{bigtab} Monte Carlo estimates for the expectation of the discrepancy statistic \eqref{metric}. Across the rows the size and subsampling rate for the series we adjust by first is kept constant. The values of $\delta$ refer to the multiples of the time step at which we keep data in the subsampling process. The values of $N$, which increase by factors of two, give the number of observations after subsampling that are used to adjust our estimate for the log-spectrum.}
\end{table}
}%

\begin{figure}
\centering
\includegraphics[width=1.5\textwidth]{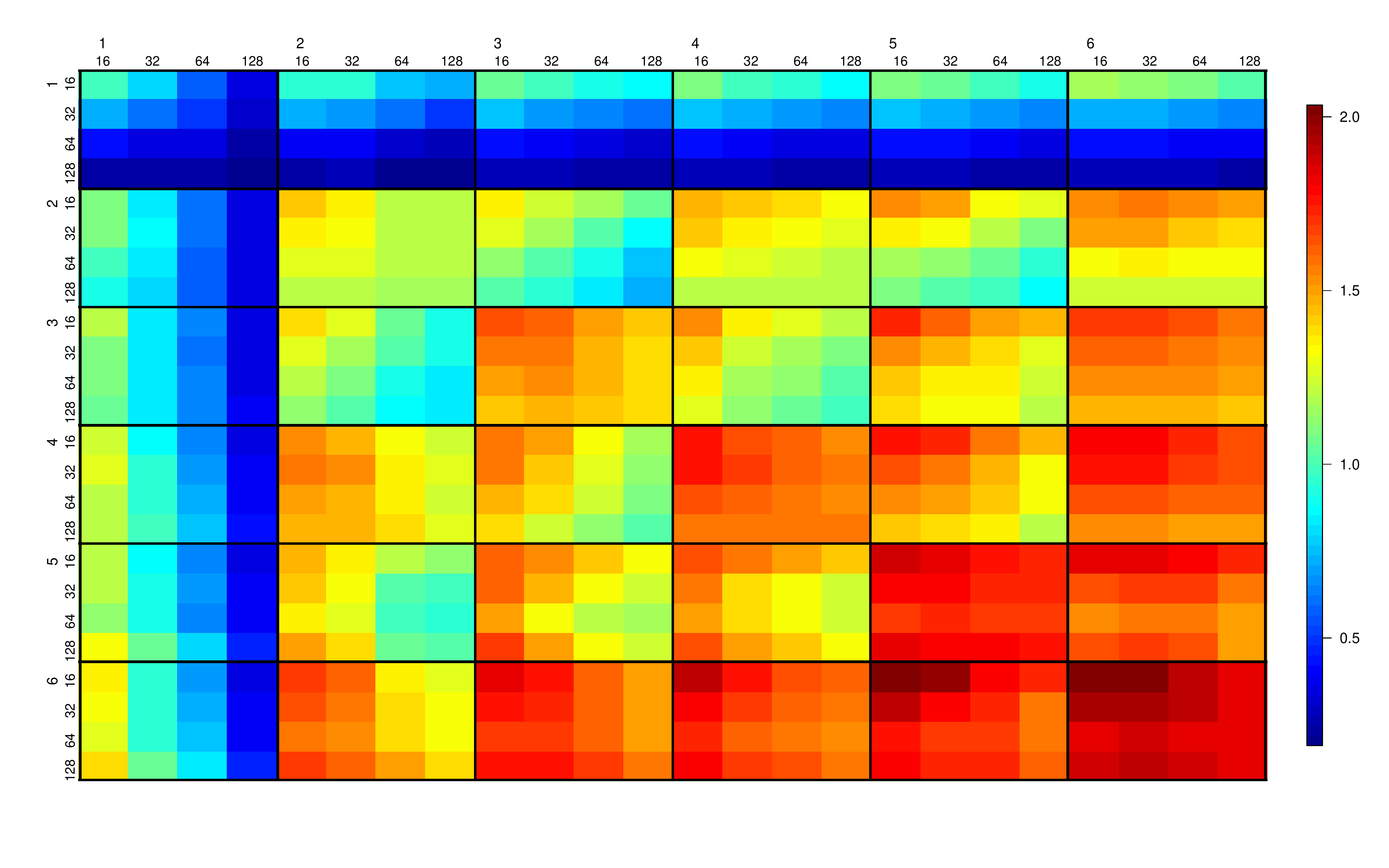}
\caption{\label{tableheatmap} A heat map diagram for quickly visually interpreting the values in Table \ref{bigtab}.}
\end{figure}

\end{landscape}

\subsection{Testing model performance given different data generating processes}
\begin{figure}
\centering
\makebox{\includegraphics[width=.48\textwidth]{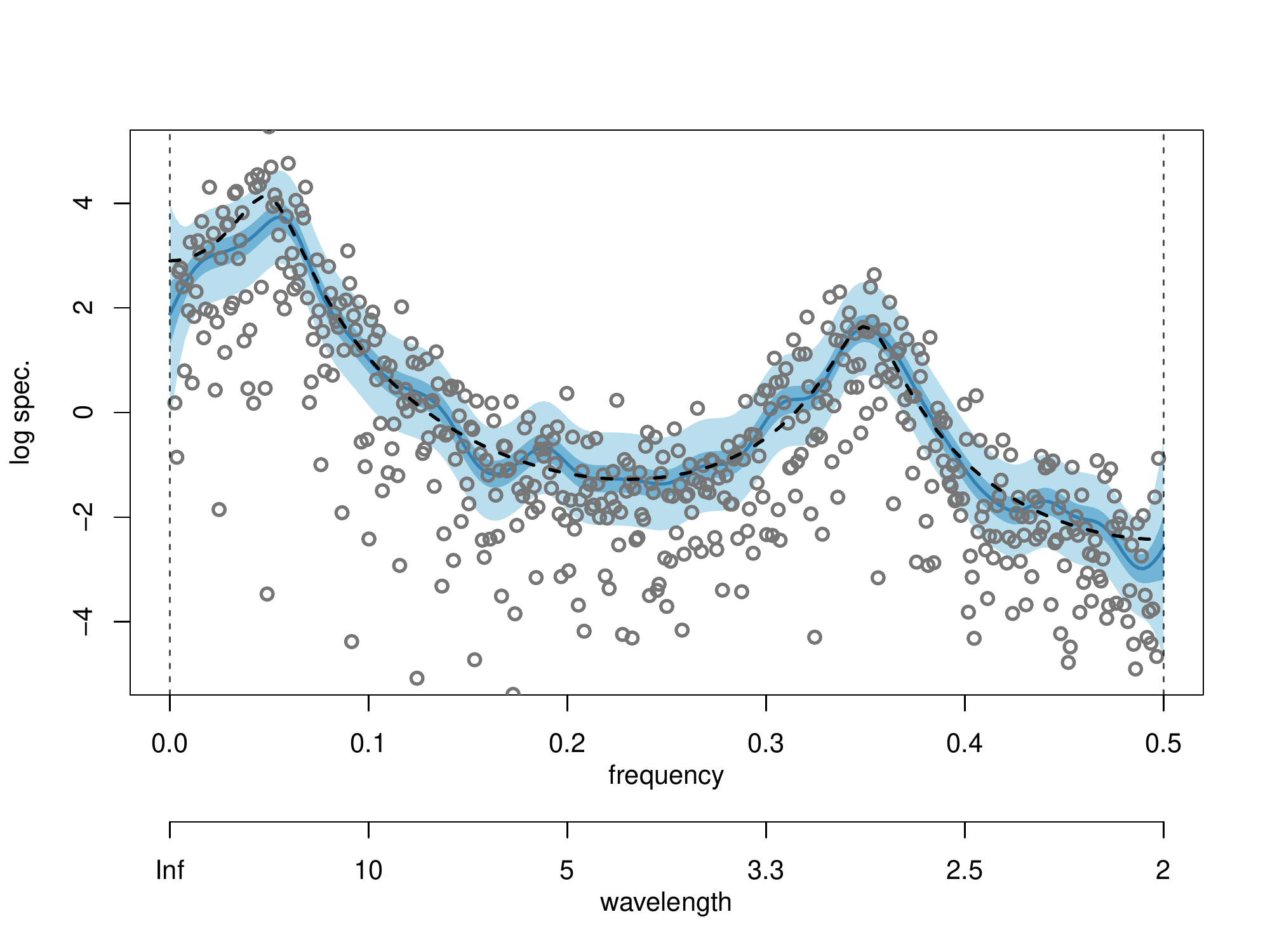}}
\makebox{\includegraphics[width=.48\textwidth]{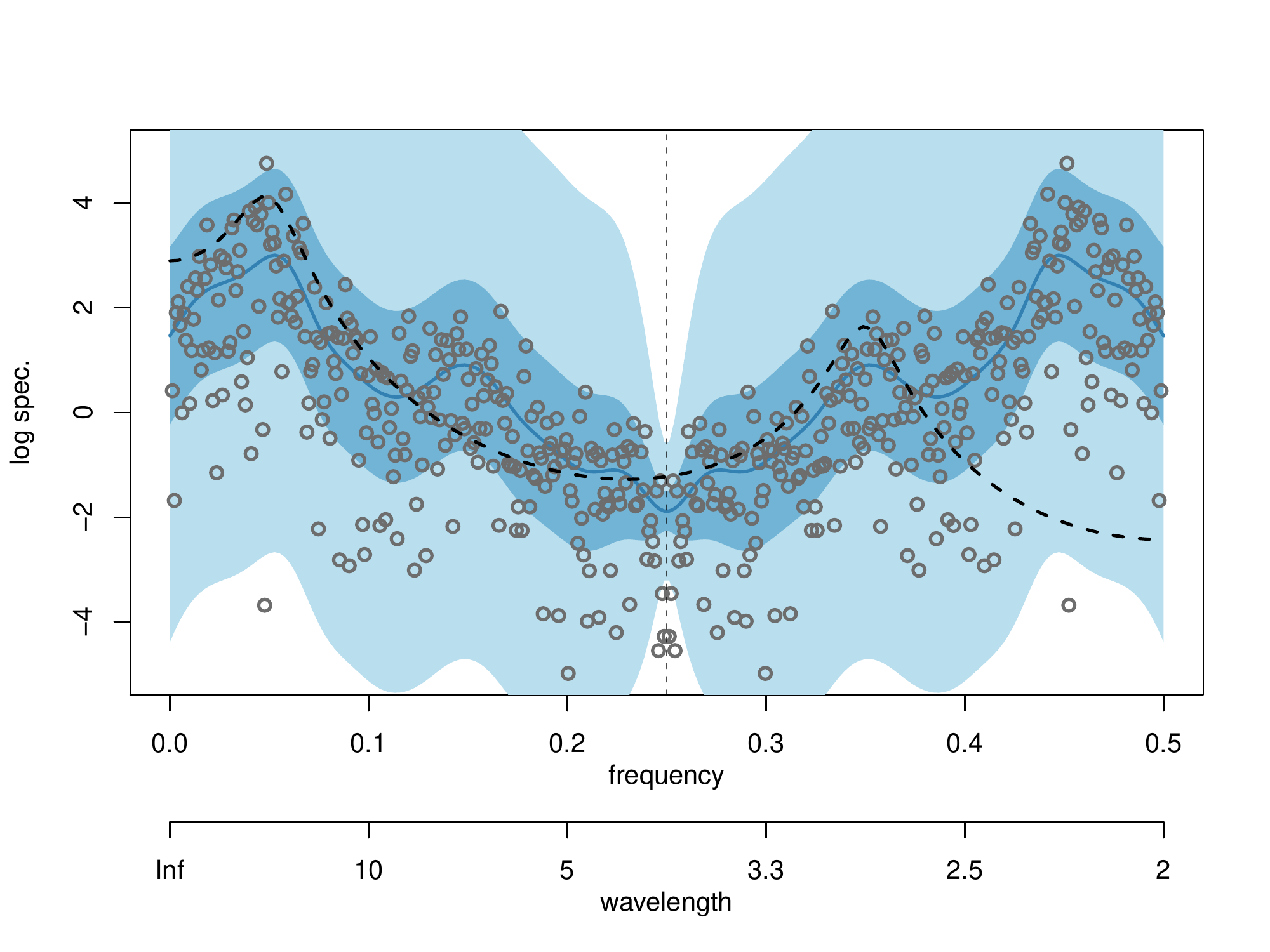}}\\
\makebox{\includegraphics[width=.48\textwidth]{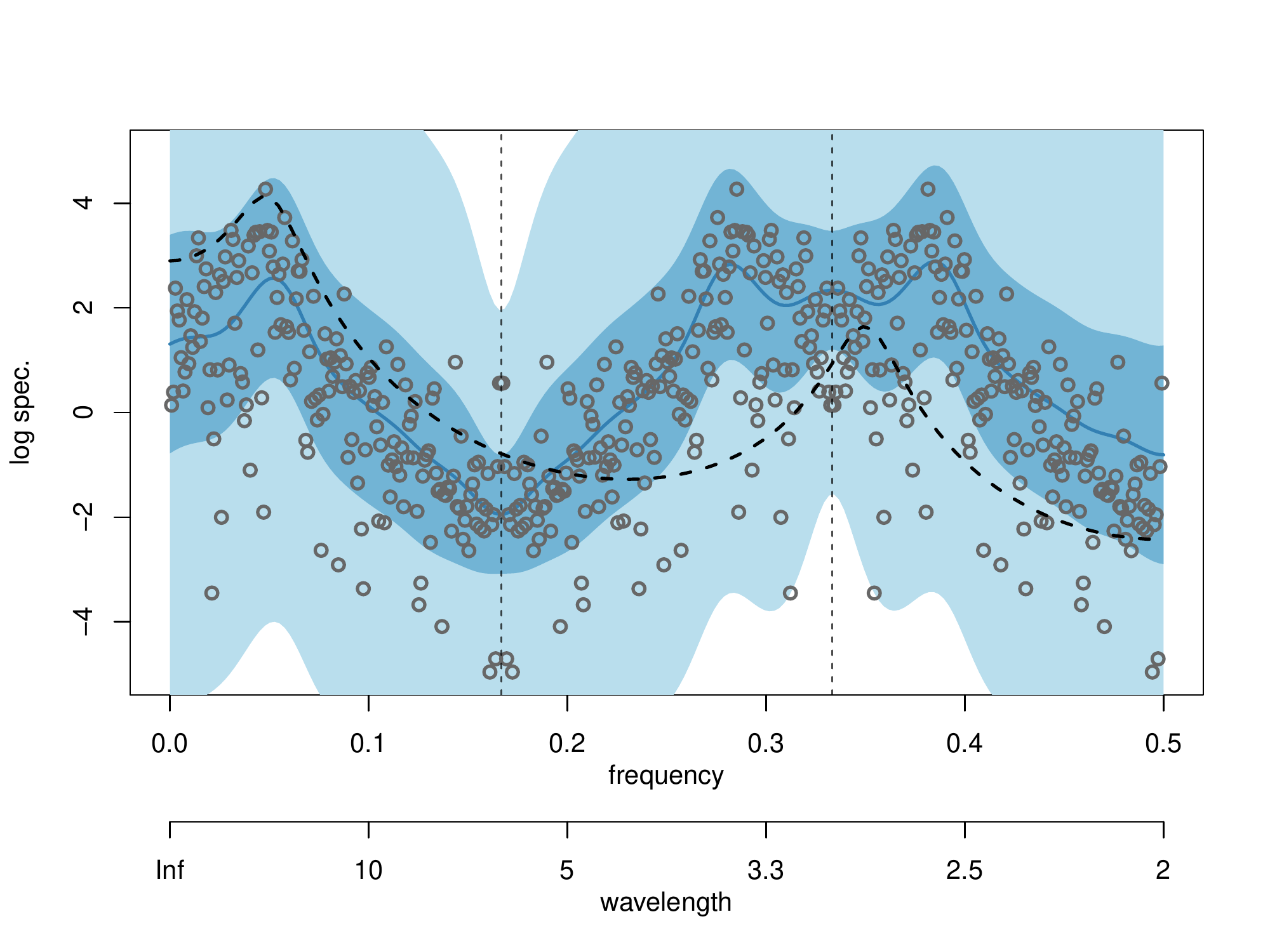}}
\makebox{\includegraphics[width=.48\textwidth]{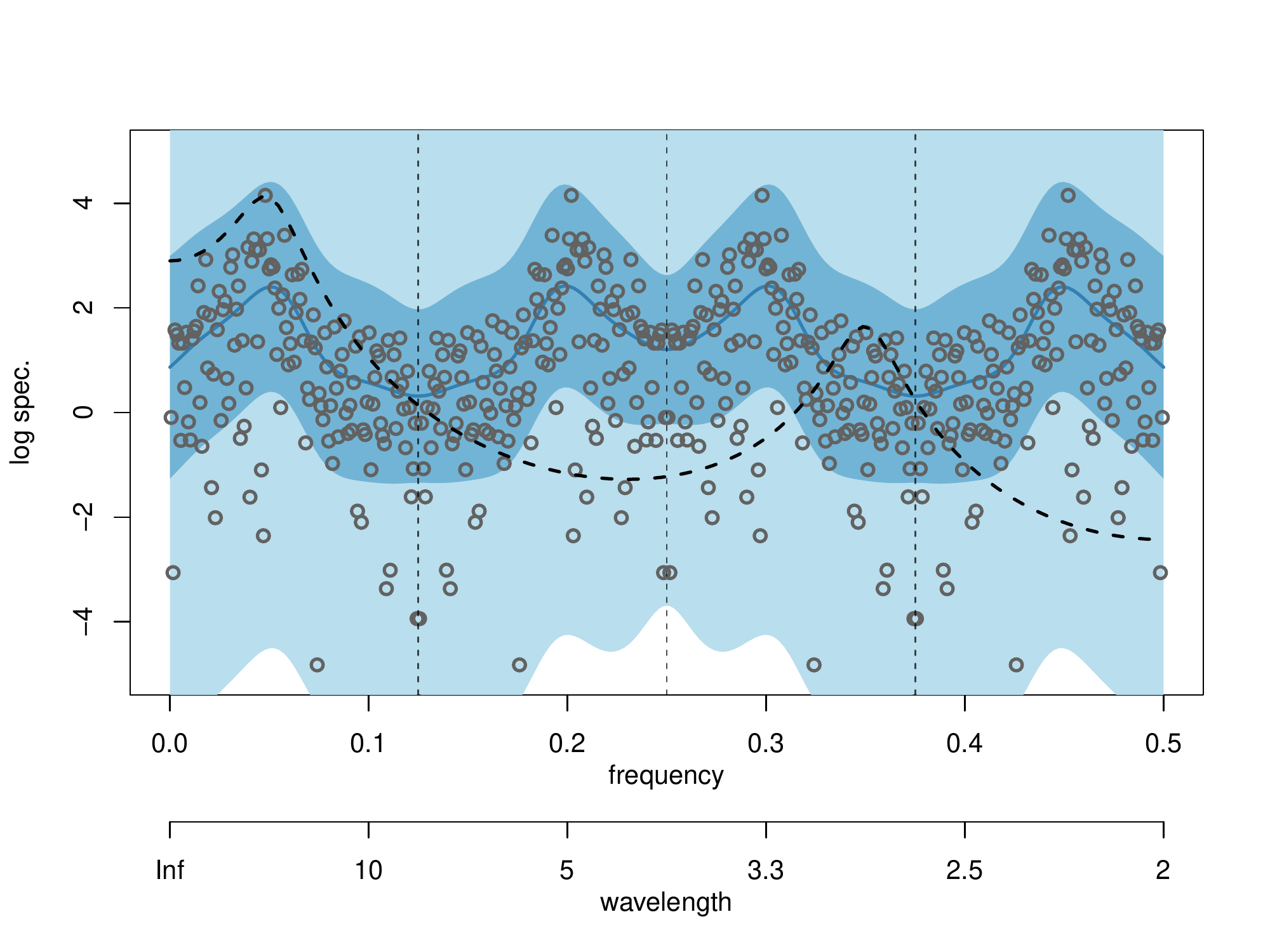}}\\
\makebox{\includegraphics[width=.48\textwidth]{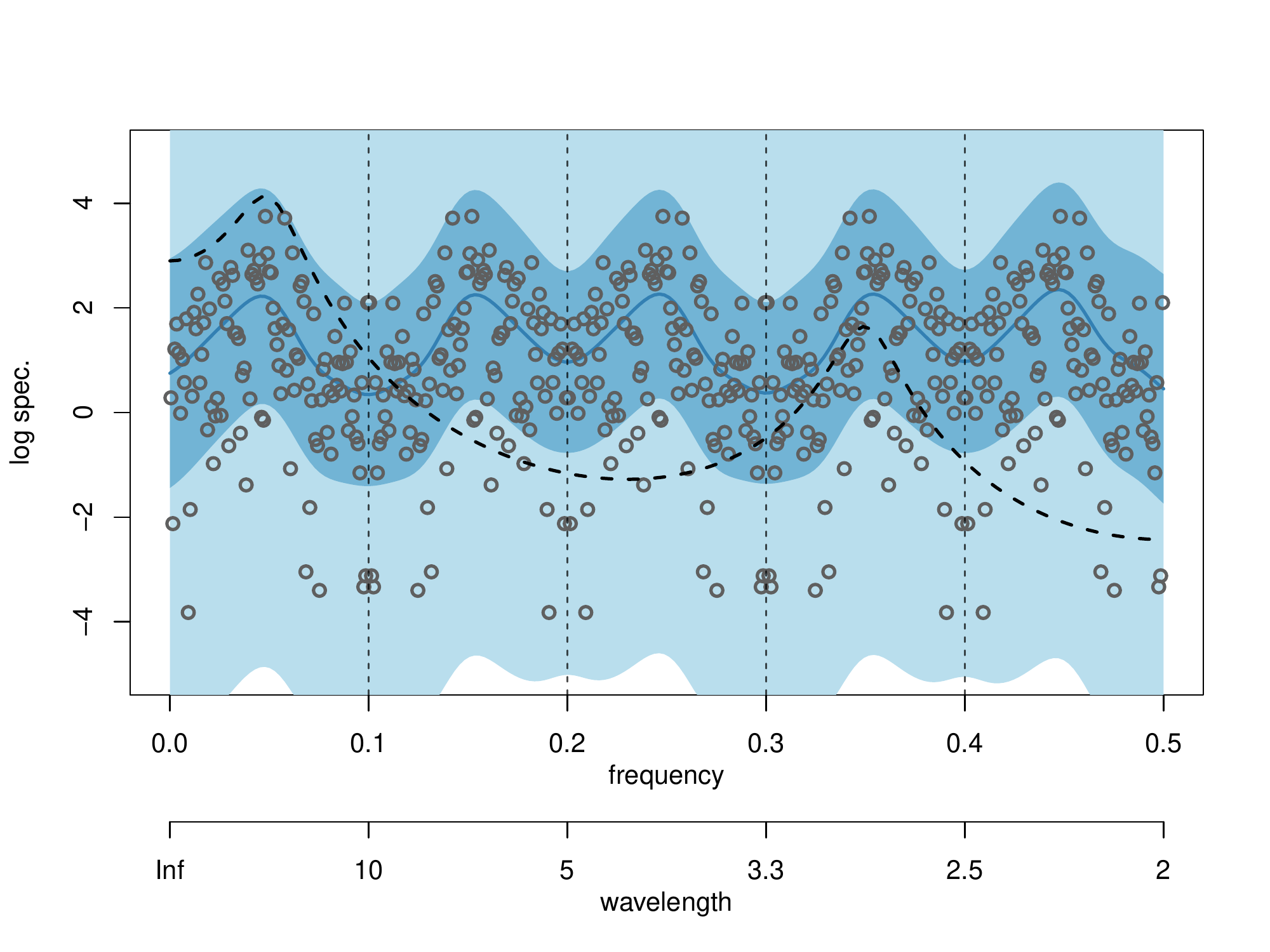}}
\makebox{\includegraphics[width=.48\textwidth]{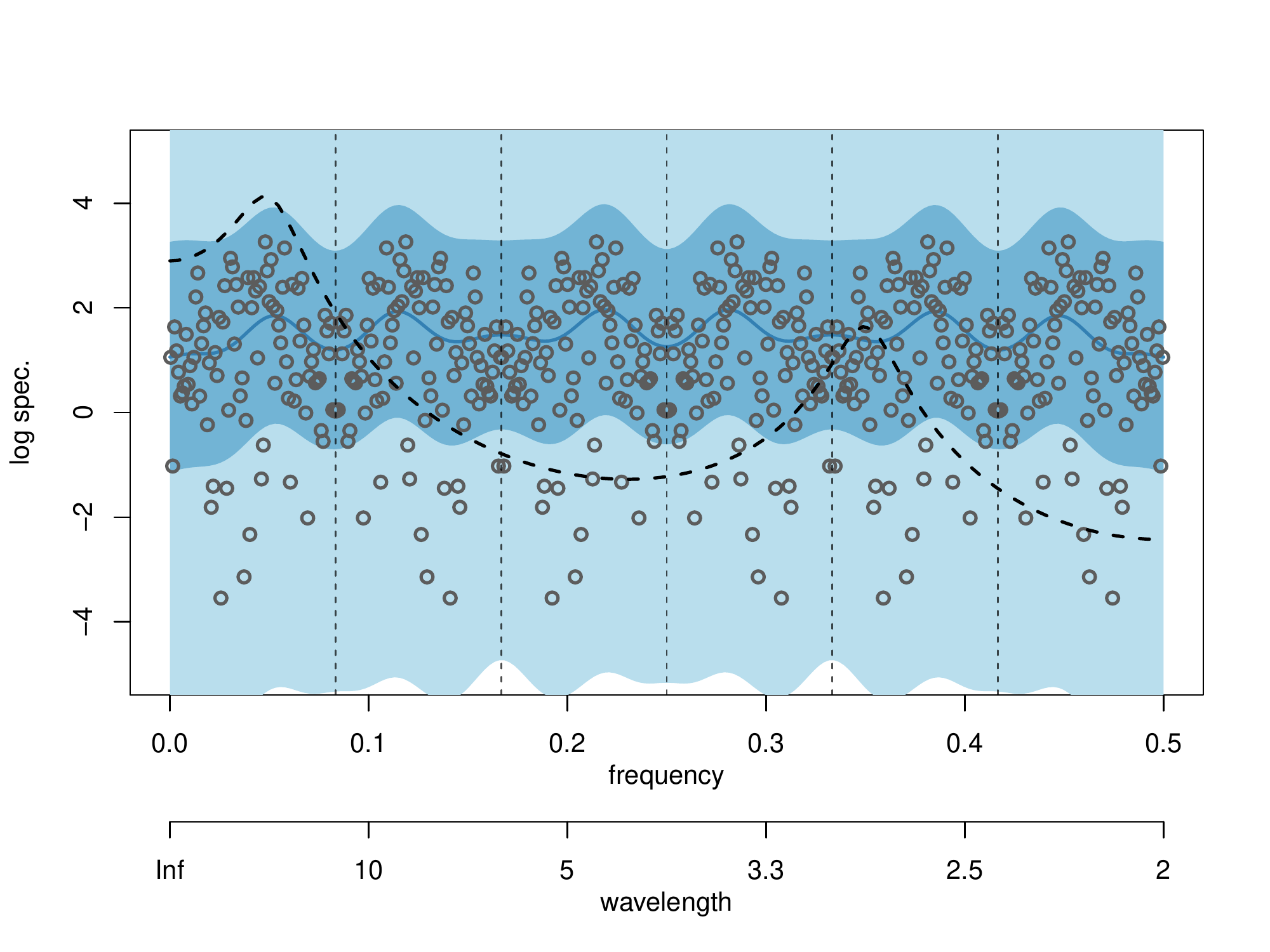}}\\
\caption{\label{dgp1} Estimates of the log-spectrum for an ARMA(4,1) model given progressively more severely subsampled versions of a simulated dataset. Moving from left to right, and top to bottom, the input data consist of 1024 data points subsampled to every first, second, third, fourth, fifth and sixth observation.}
\end{figure}

\begin{figure}
\centering
\makebox{\includegraphics[width=.48\textwidth]{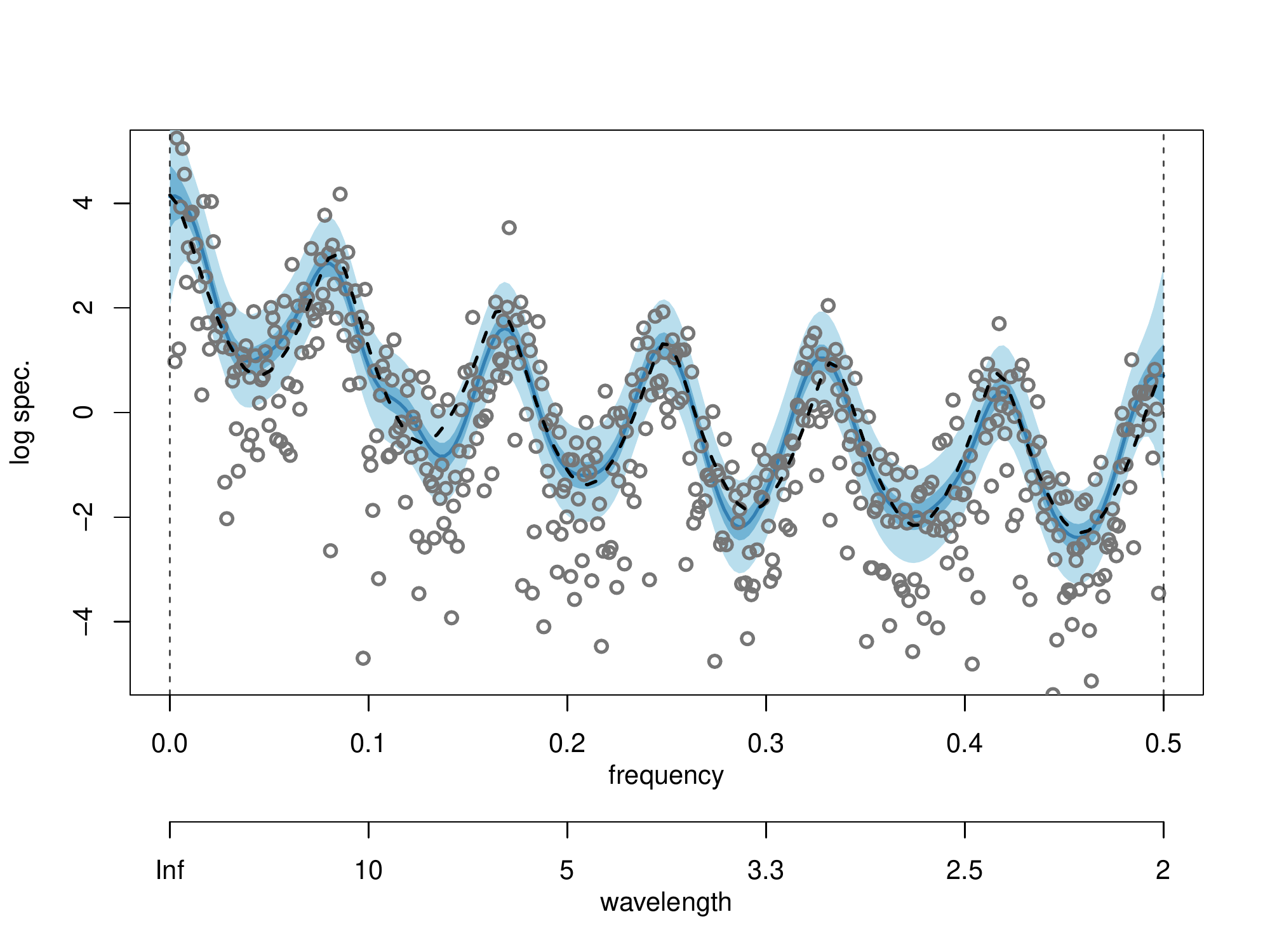}}
\makebox{\includegraphics[width=.48\textwidth]{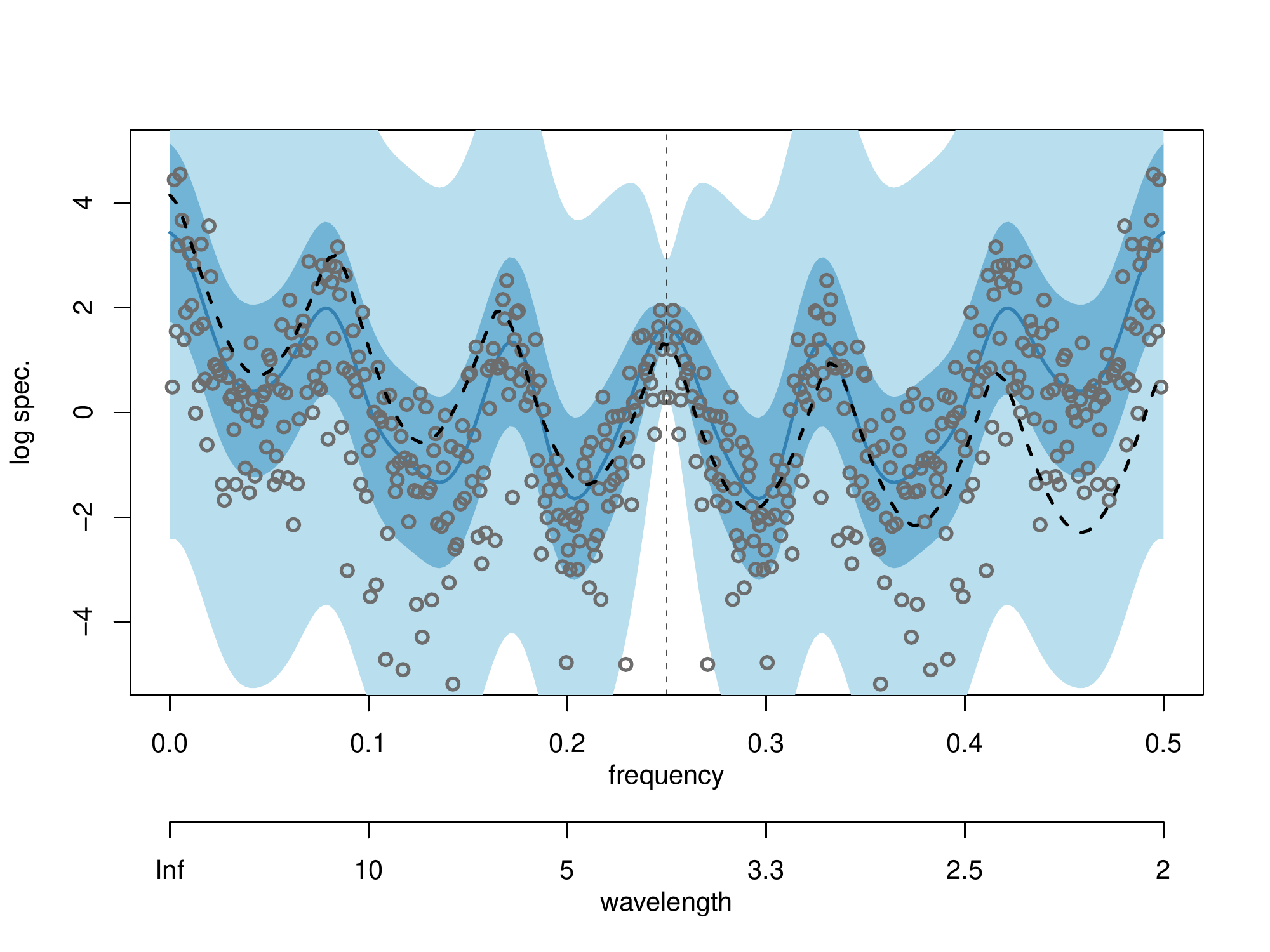}}\\
\makebox{\includegraphics[width=.48\textwidth]{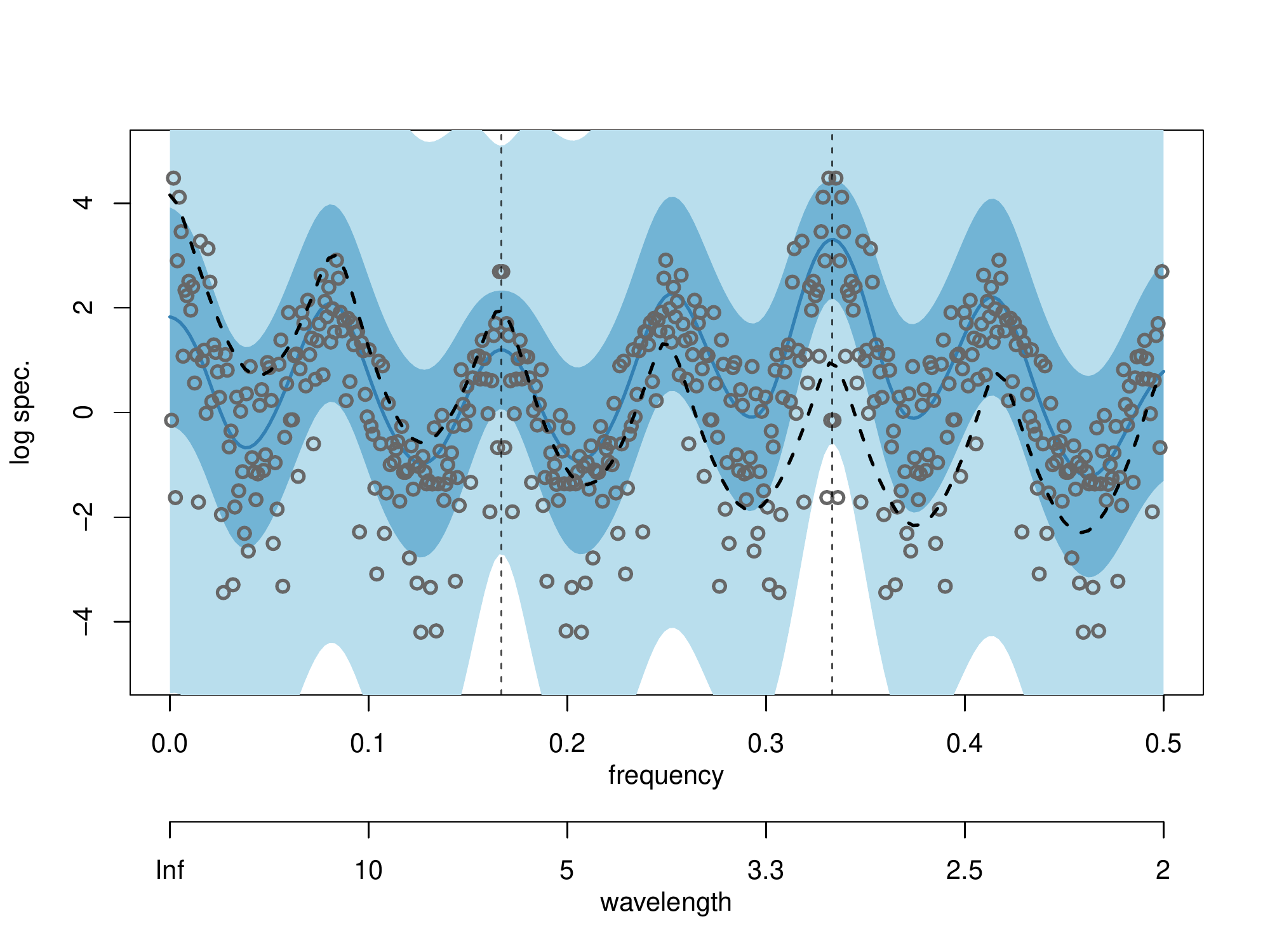}}
\makebox{\includegraphics[width=.48\textwidth]{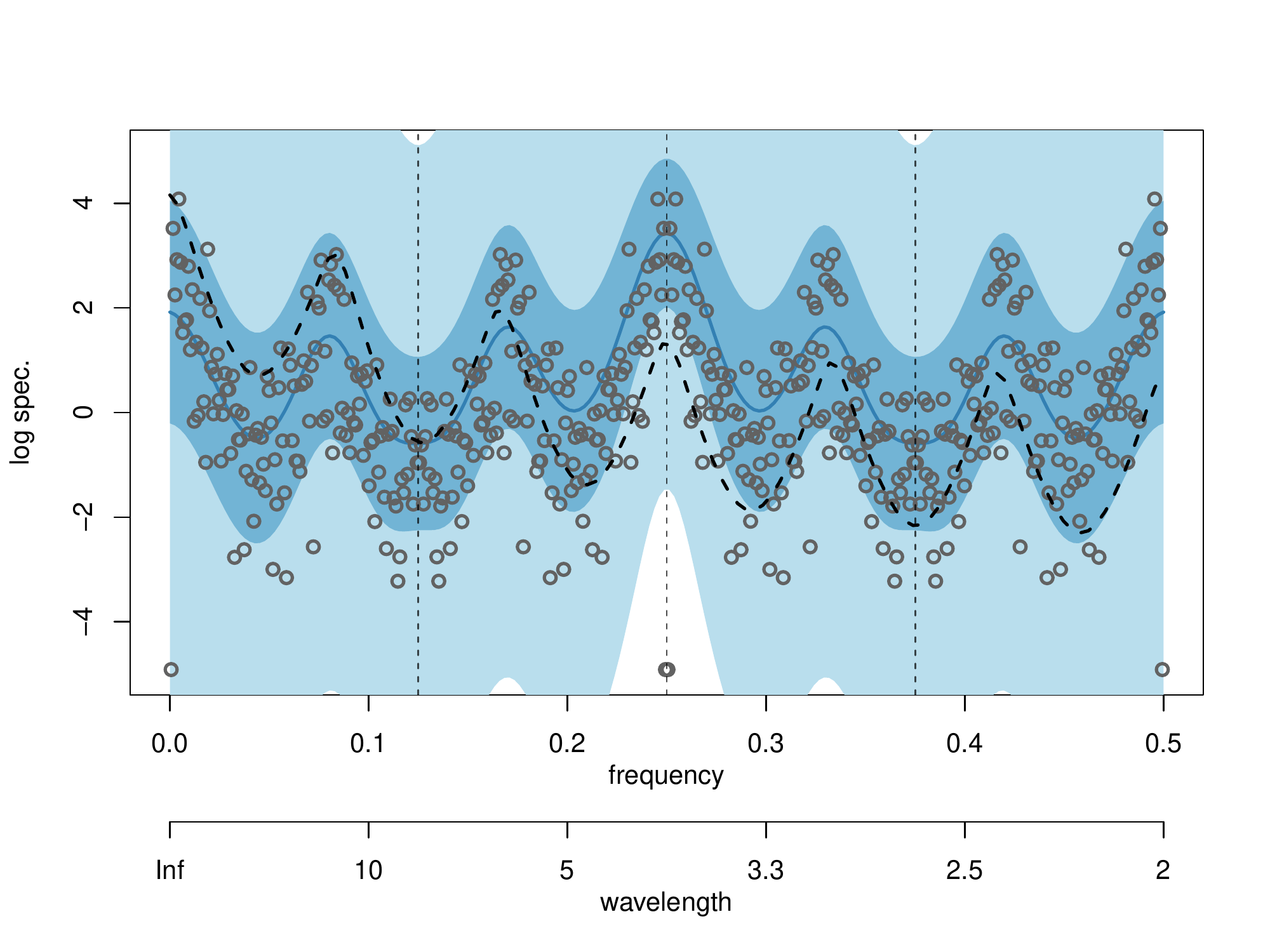}}\\
\makebox{\includegraphics[width=.48\textwidth]{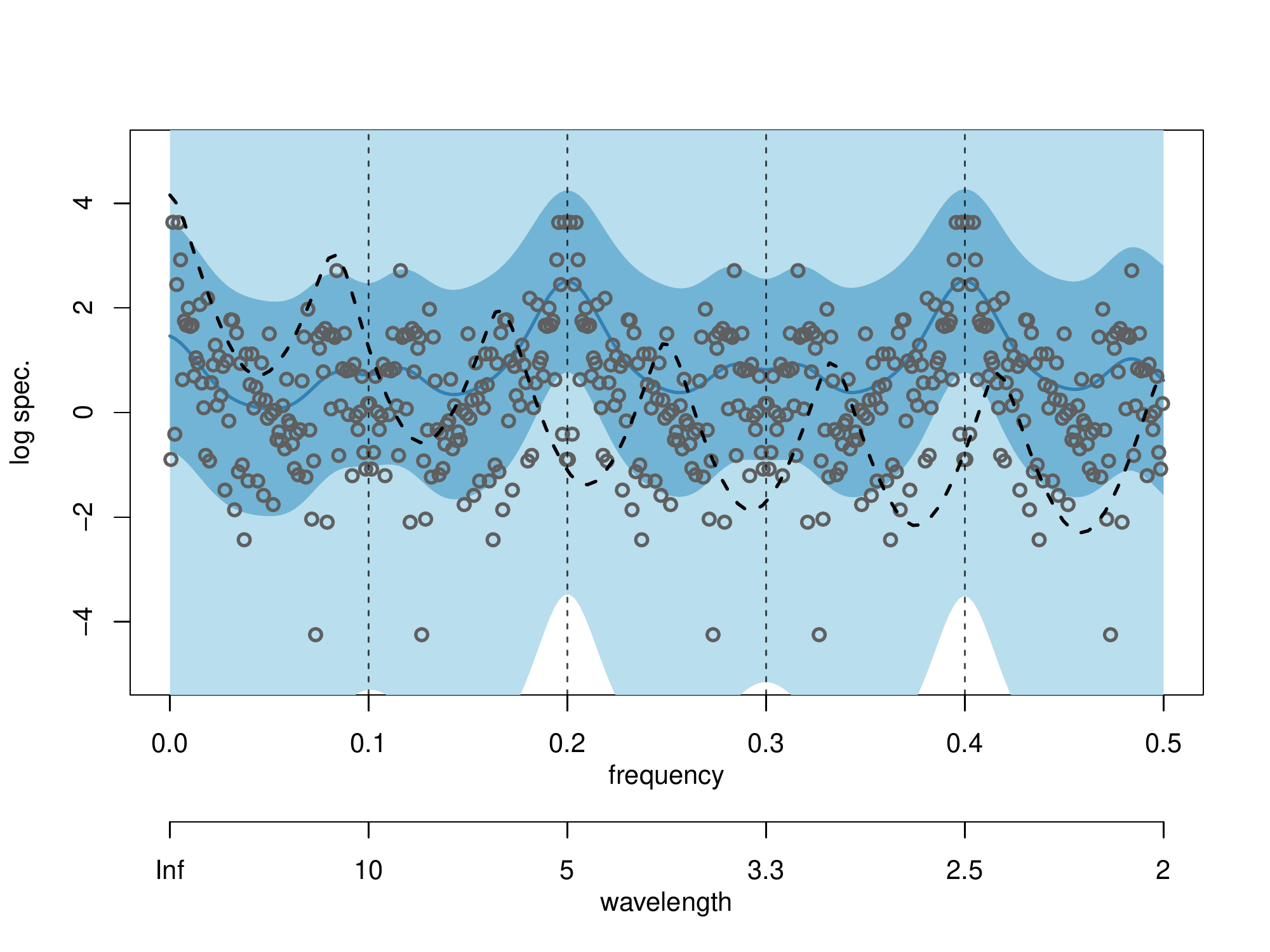}}
\makebox{\includegraphics[width=.48\textwidth]{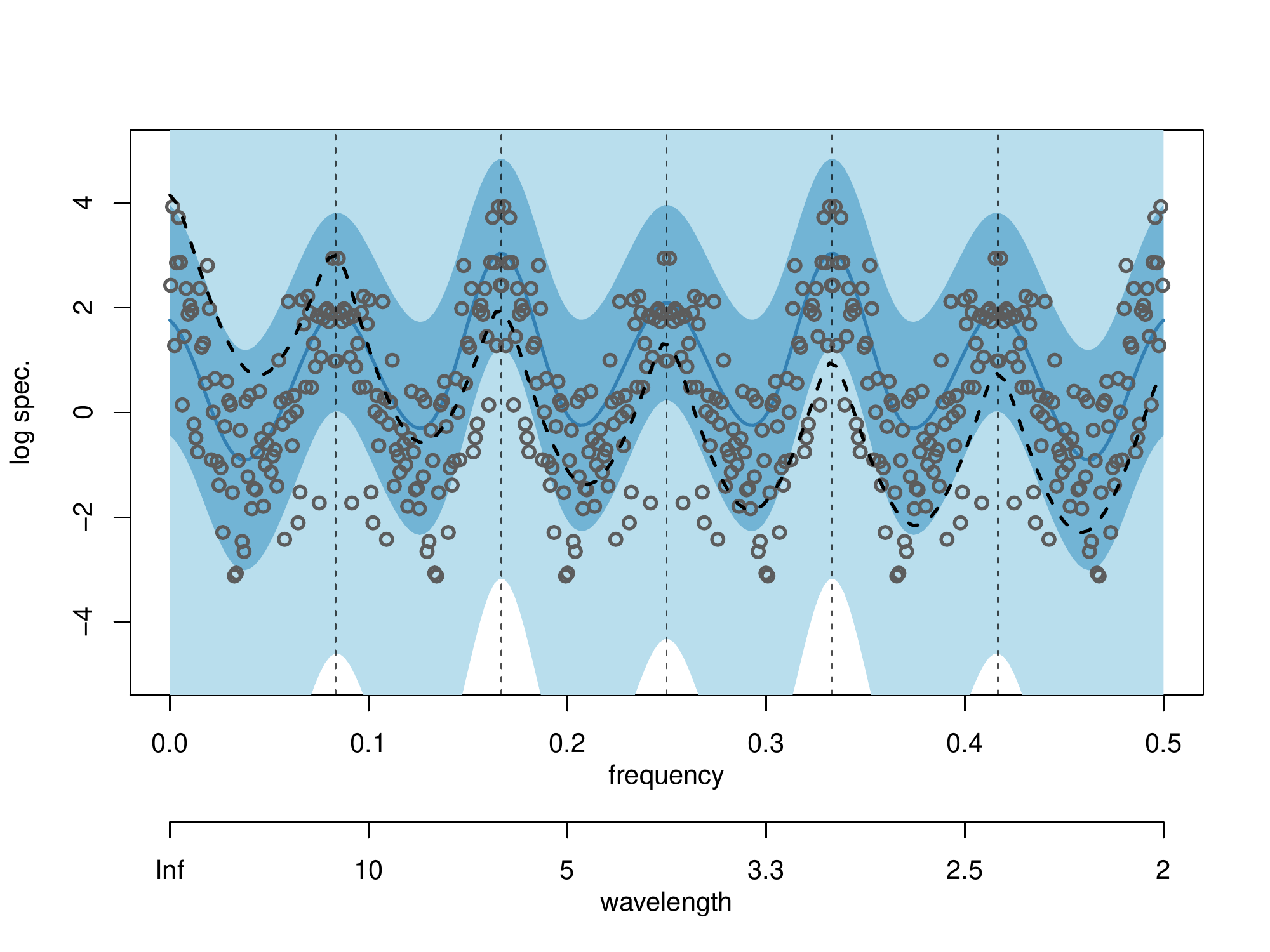}}\\
\caption{\label{dgp2}Estimates of the log-spectrum for an SARMA$(0,0)(1,1)_{12}$ model given progressively more severely subsampled versions of a simulated dataset. Moving from left to right, and top to bottom, the input data consist of 1024 data points subsampled to every first, second, third, fourth, fifth and sixth observation.}
\end{figure}

\begin{figure}
\centering
\makebox{\includegraphics[width=.48\textwidth]{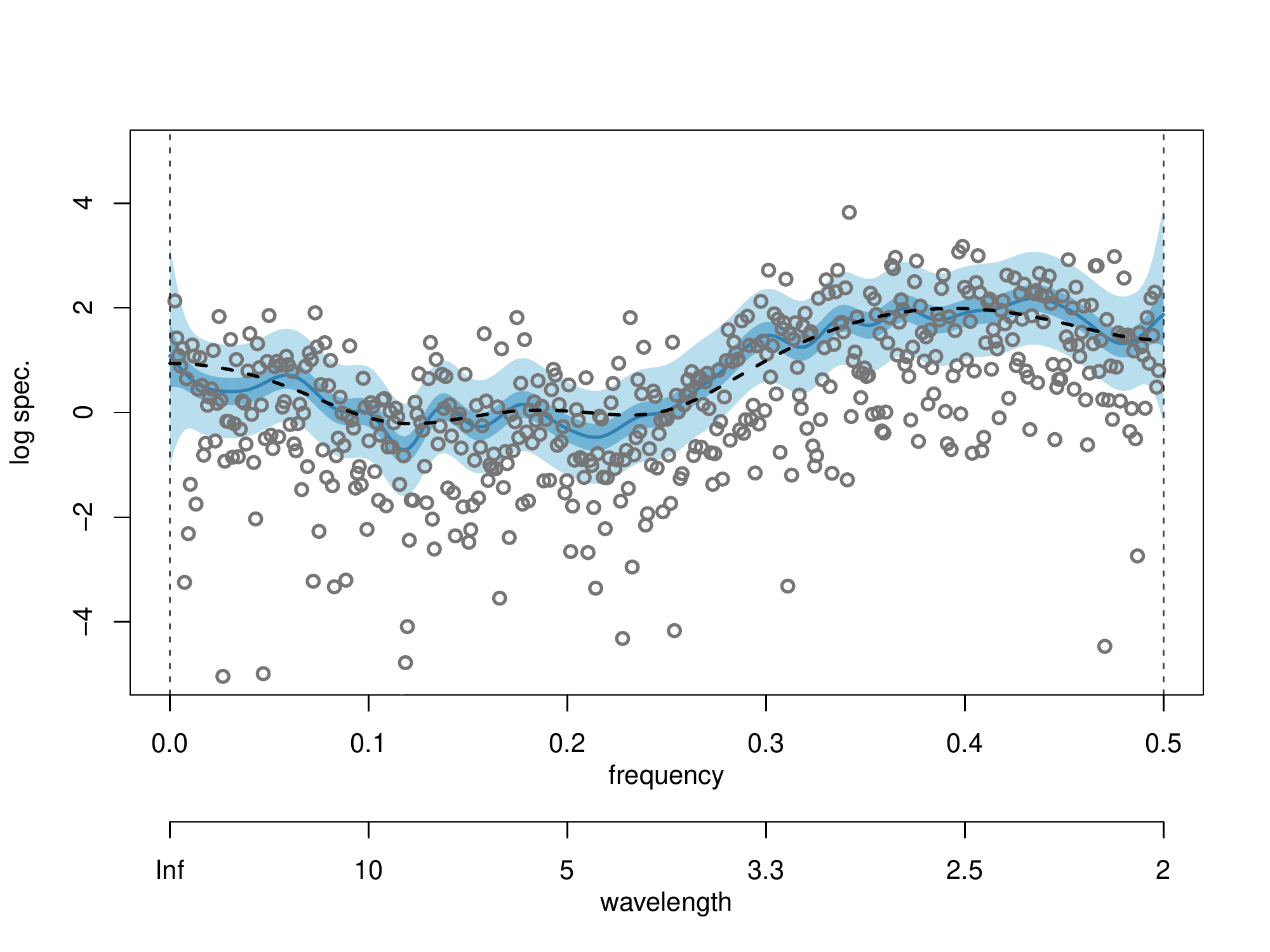}}
\makebox{\includegraphics[width=.48\textwidth]{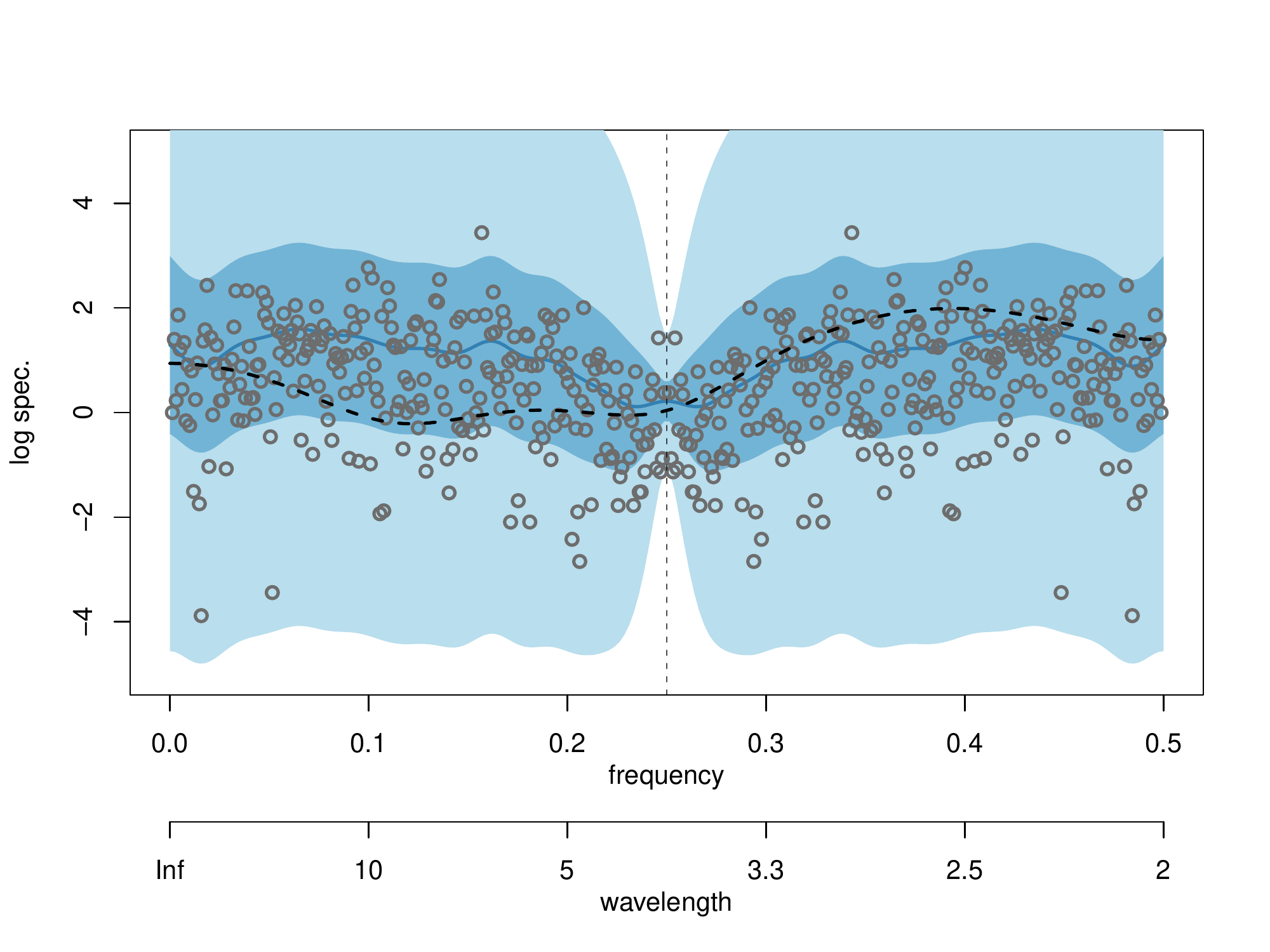}}\\
\makebox{\includegraphics[width=.48\textwidth]{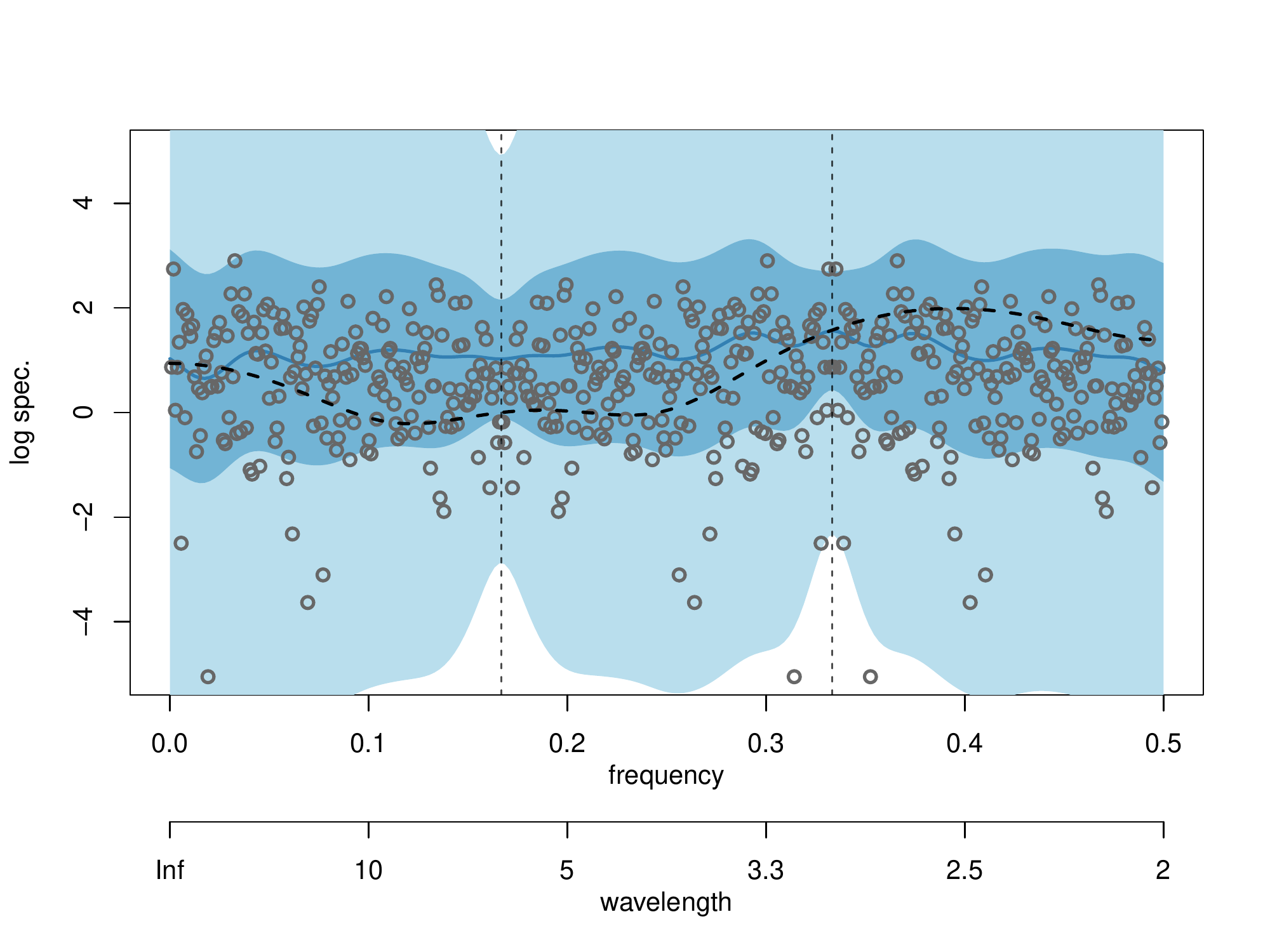}}
\makebox{\includegraphics[width=.48\textwidth]{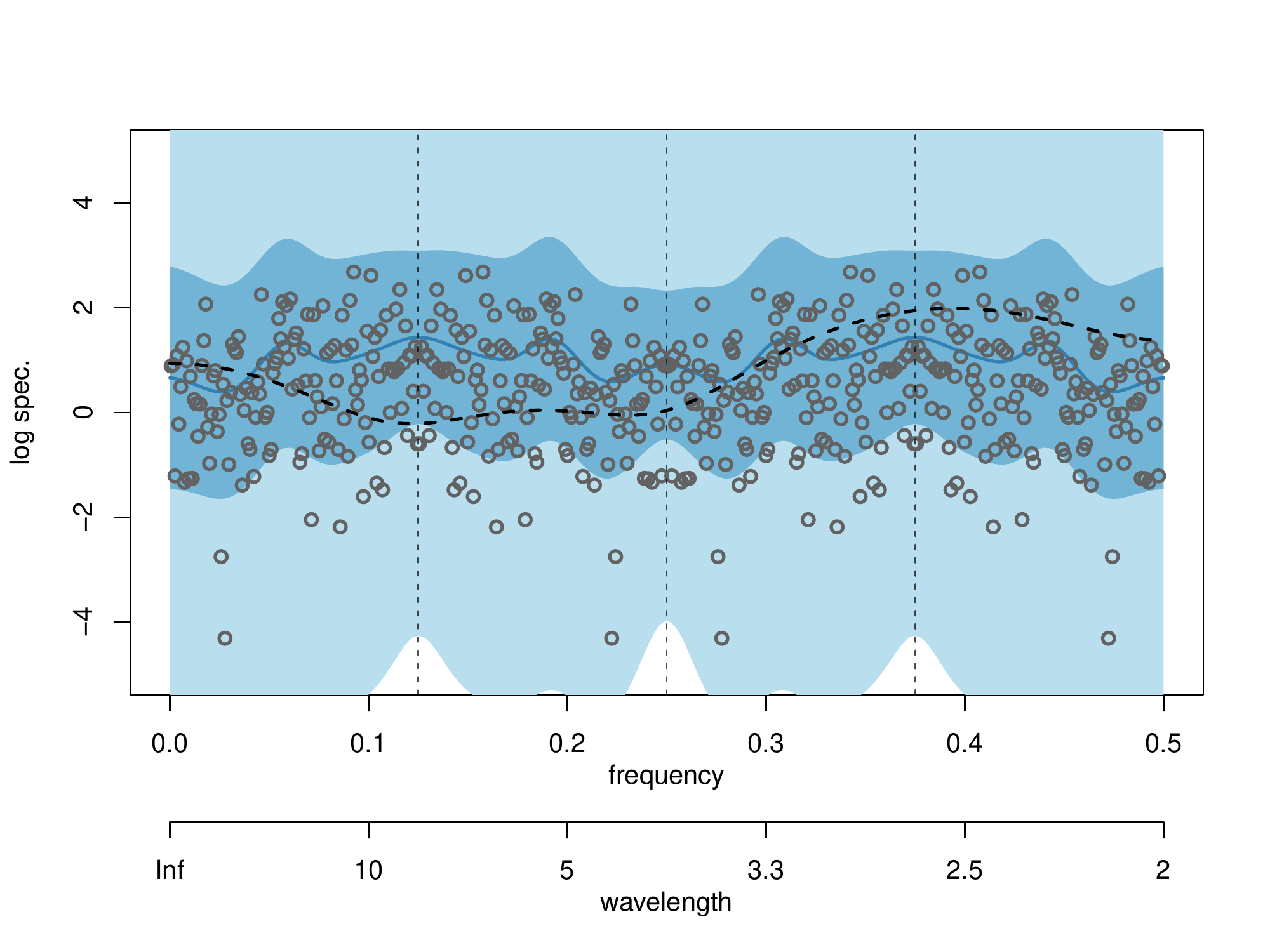}}\\
\makebox{\includegraphics[width=.48\textwidth]{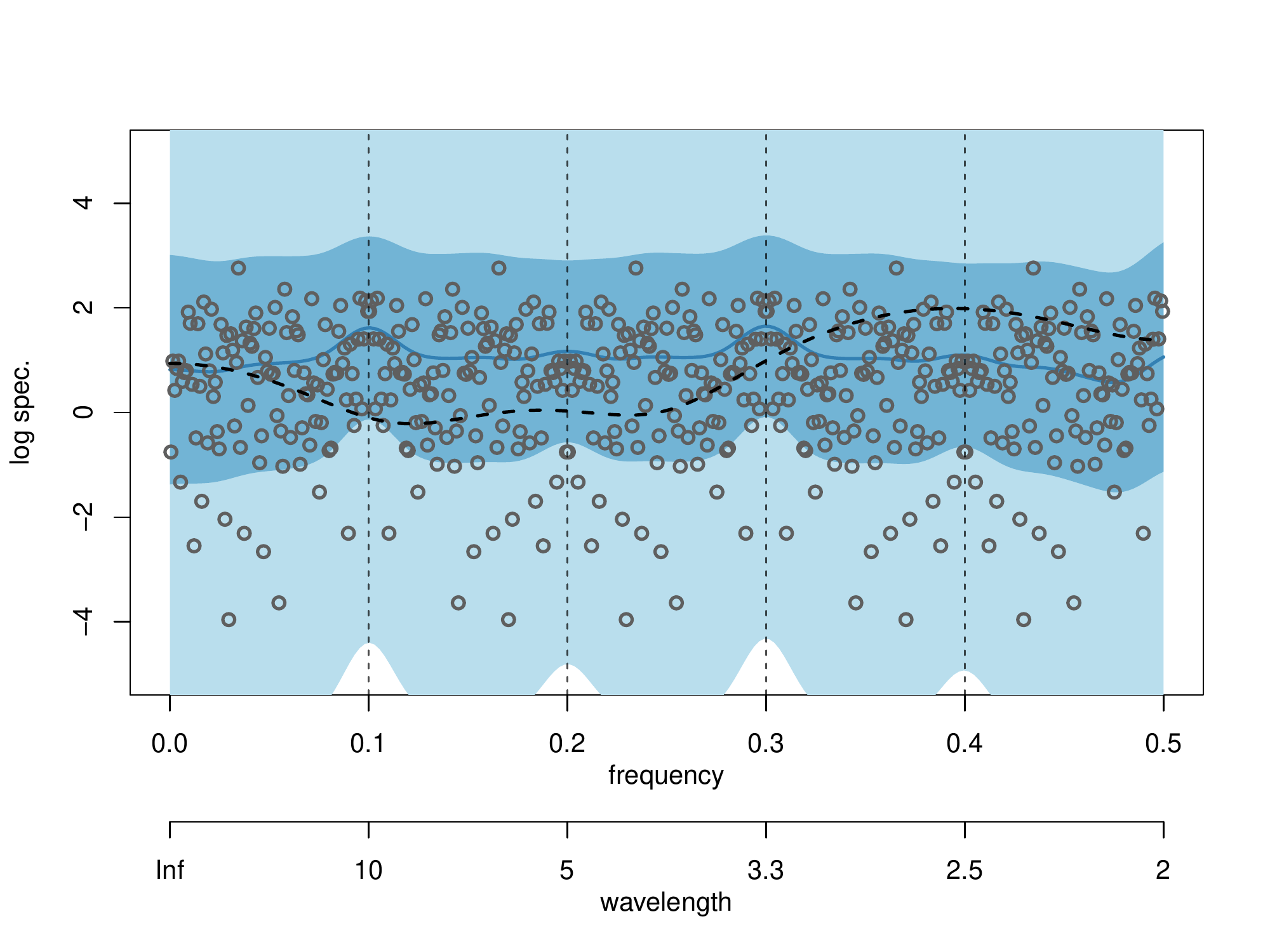}}
\makebox{\includegraphics[width=.48\textwidth]{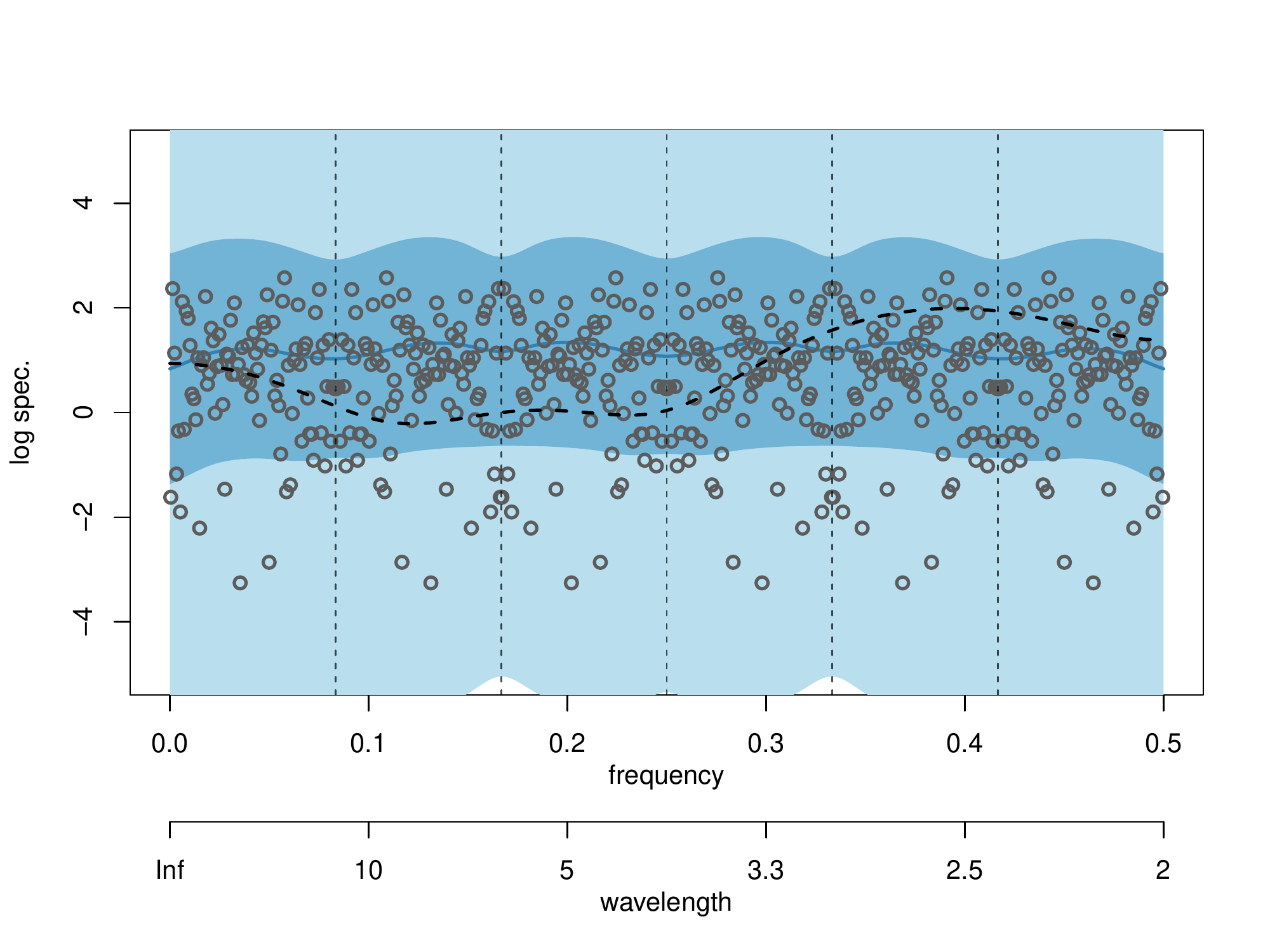}}\\
\caption{\label{dgp3}Estimates of the log-spectrum for an ARMA(0,5) model given progressively more severely subsampled versions of a simulated dataset. Moving from left to right, and top to bottom, the input data consist of 1024 data points subsampled to every first, second, third, fourth, fifth and sixth observation.}
\end{figure}

We now take a look at some example log-spectrum inferences given data simulated from a range of SARMA models. The SARMA models include an ARMA(4,1) model with two distinct sharp spectral peaks, a SARMA$(0,0)(1,1)_{12}$ model with 7 regularly-spaced peaks, and a ARMA(0,5) model with a very smooth spectrum. Note that although all these example models are of the SARMA form, our model for the log-spectrum is in no way restricted to, or especially tuned to accommodate data from such models. The choice to focus on these examples in particular arose from the ease with which we may simulate values from them, and from their familiarity amongst the time series community. Plots communicating the Bayes linear adjusted expectations, and approximate conservative credible regions, for the log-spectra given data sub-sampled at different rates are presented in Figures \ref{dgp1}--\ref{dgp3}. The figures shed more light on the way the aliasing phenomenon effects our ability to infer a process' spectrum. Our model, having been 
formulated in anticipation of the aliasing phenomenon, is able to recognise that more severely sub-sampled data leave us with greater uncertainty for the spectrum, and, more significantly, with uncertainty that obeys a very specific symmetry pattern.

\subsection{Testing model performance against simpler interpolation-based methods}
In this subsection we focus on the comparison of our proposed method for log-spectrum estimation with a much simpler method, according to which data are interpolated to fill in missing values before conventional (constant sampling rate) spectral analysis tools are employed. We focus on this particular alternative to our own method, having been made aware that it represents a temptingly fast and convenient way to produce a spectral estimate.

As we have established in \cite{NPES15} and reiterated in Section \ref{supplementarytheory} of this document, the aliasing phenomenon means that subsampled data leave an unresolvable symmetry in the range of plausible spectra that characterize a stationary process. By interpolating observed data to fill in the missing values, we disregard the ambiguity for the spectrum and, in fact, effectively supplant it with information biasing the spectral power towards the lowest frequencies since the majority of interpolation routines automatically calculate the least rough or oscillatory curve passing through the observed data.

This problem is exemplified in the following example, although it can be expected to appear to a less significant effect for all subsampled series whose spectra contain power above the Nyquist frequency associated with the lower sampling rate.

\newpage

\begin{example}
In Figure \ref{interpseries} we plot a series of data simulated from an AR(2) process with a spectral peak at frequency $0.35$. Five sixths of the original series are designated the historical series and are subsampled to leave only every other observation. The retained observations are then passed to a routine performing cubic spline interpolation, producing an interpolated series with which we can perform conventional spectral analyses.

With Figure \ref{regspecrighta}, we provide an impression of the type of symmetric uncertainty left unresolved by the subsampled data. Figure \ref{arspec} then shows two out-of-the-box estimates of the log-spectrum based on the interpolated data. We can quite easily see here that the interpolation routine has, in effect, steered us towards the spectrum with most power at the lower frequencies that is still consistent with subsampled data. In Figure \ref{regspecwrong}, we plot the adjusted expectation as calculated using our model and the same interpolated dataset, and observe the same problem. In Figure \ref{regspecright}, we plot the model's expectation adjusted by the two portions of the observed data (and only the observed data) in accordance with the theory described in \cite{NPES15}, and see that the bias in the spectrum towards low frequencies is removed.

\begin{figure}
\centering
\includegraphics[width=.6\textwidth]{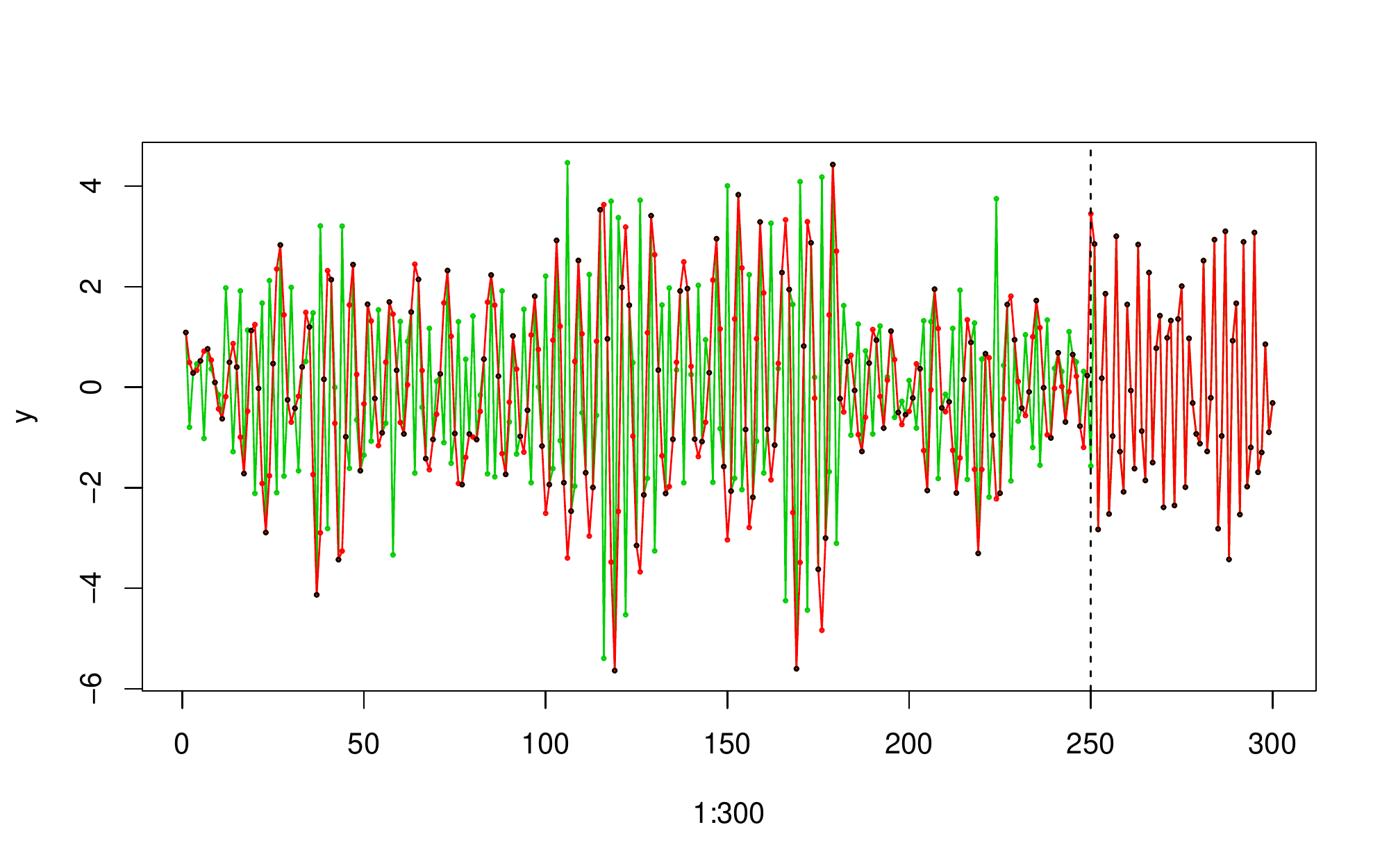}
\caption{\label{interpseries} Plotted in black are the example data considered to be observable for the test of interpolation-based spectrum estimation. The interpolated data are plotted in red, while the true unobserved values are plotted in green.}
\end{figure}

\begin{figure}
\centering
\includegraphics[width=.6\textwidth]{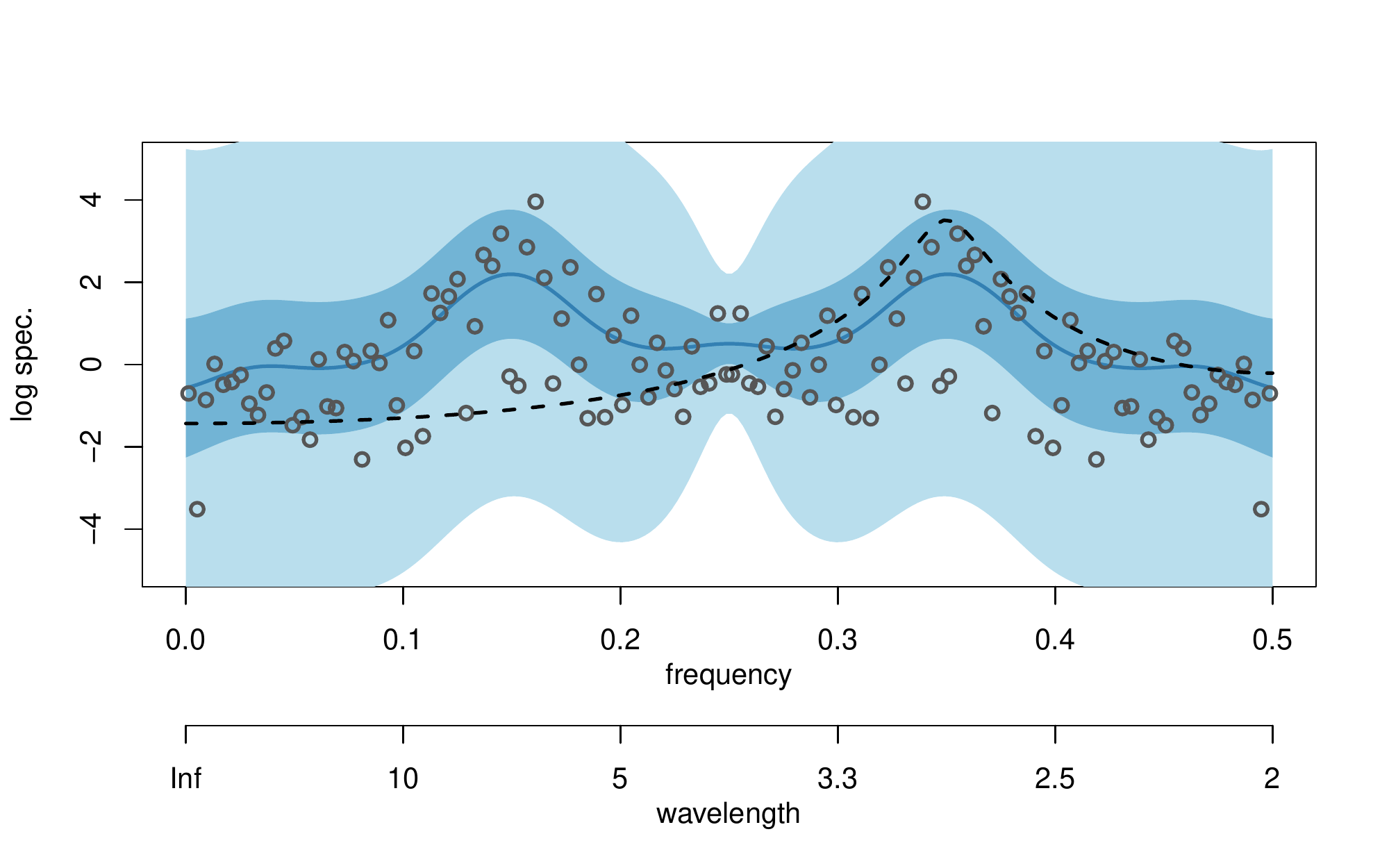}
\caption{\label{regspecrighta} A plot of the adjusted log-spectrum according to the linear Bayes model given just the subsampled data.}
\end{figure}

\begin{figure}
\centering
\includegraphics[width=.6\textwidth]{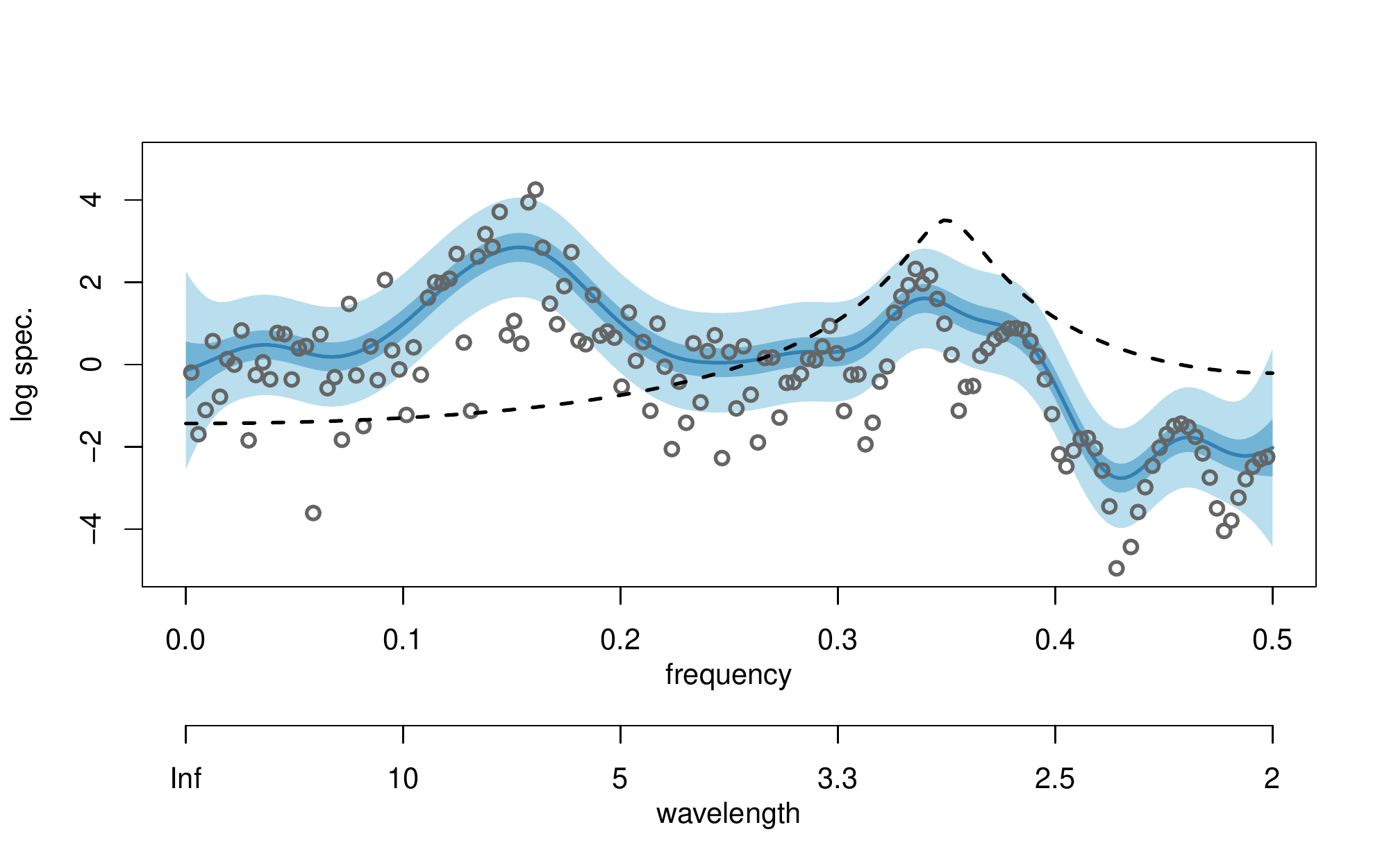}
\caption{\label{regspecwrong} A plot of the adjusted log-spectrum according to the linear Bayes model, given the observed data and interpolated data as though they had both been observed. The grey circles here mark the values of the log-periodogram values calculated from the interpolated dataset.}
\end{figure}

\begin{figure}
\centering
\includegraphics[width=.6\textwidth]{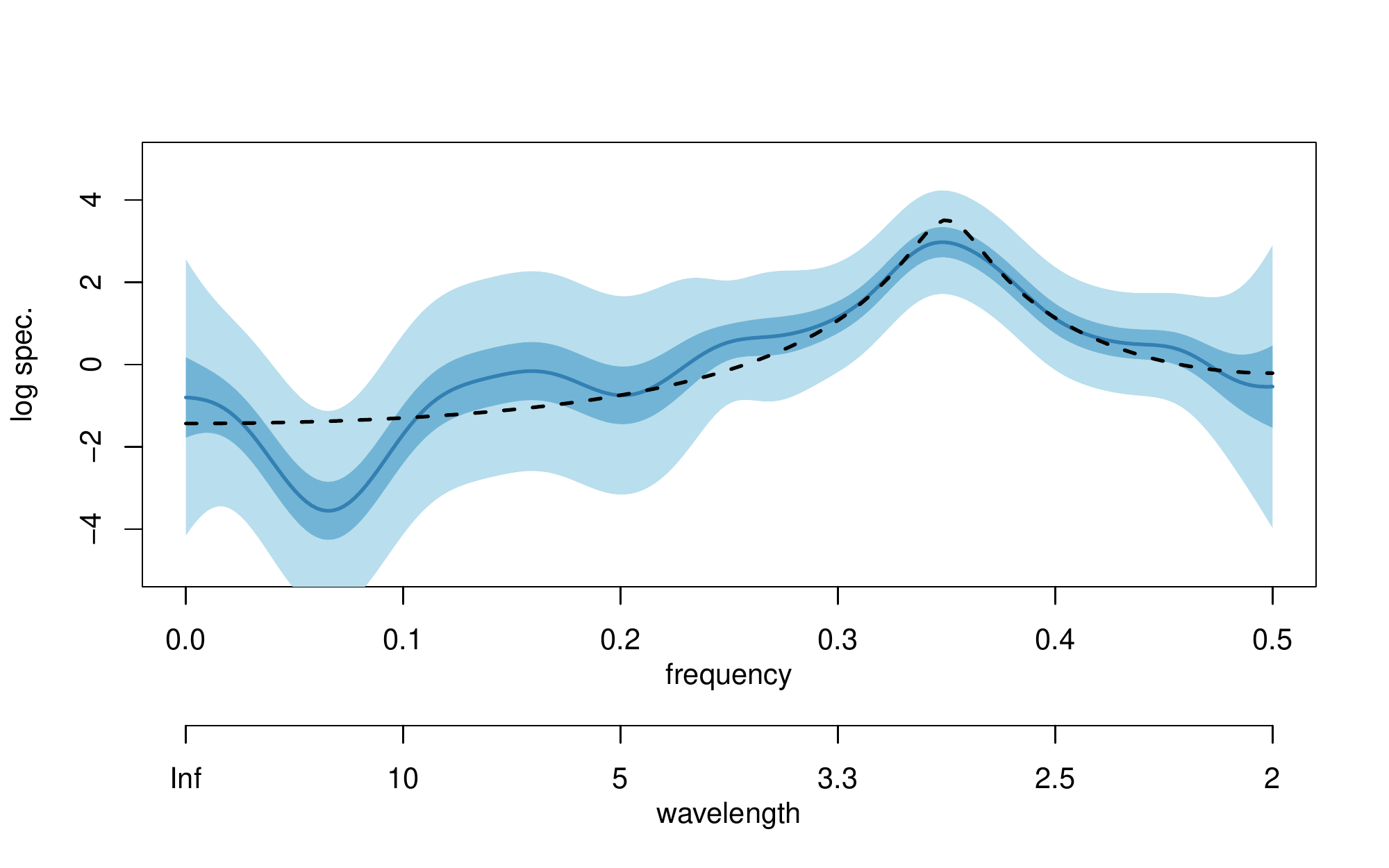}
\caption{\label{regspecright} A plot of the adjusted log-spectrum according to the linear Bayes model given just the observed data.}
\end{figure}

\begin{figure}
\centering
\includegraphics[width=.6\textwidth]{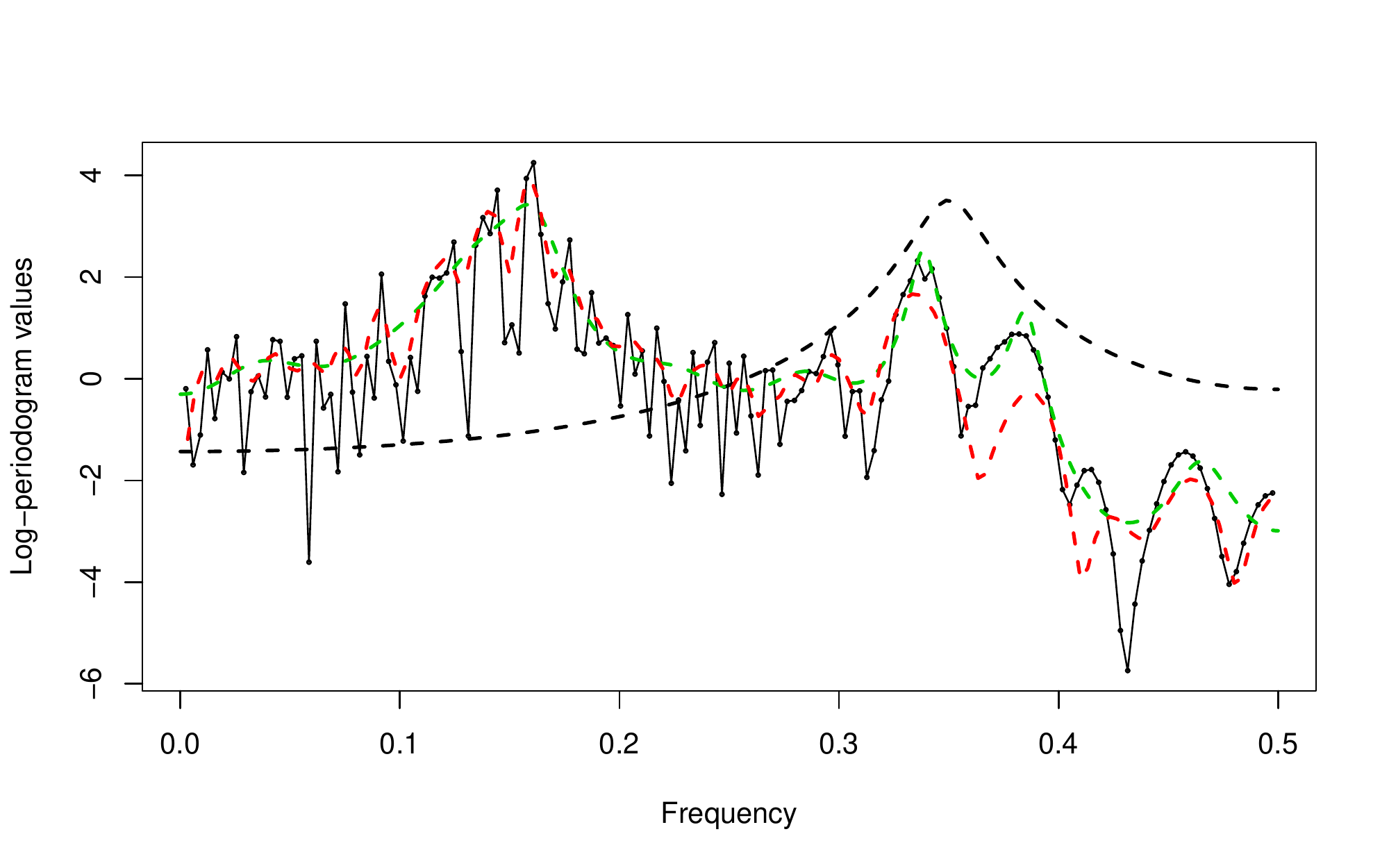}
\caption{\label{arspec} A plot of log-spectrum estimates calculated using the AR fitting method and the smoothed periodogram method built into the R \texttt{stats} package, given the observed data and interpolated data as though they had both been observed.}
\end{figure}
\end{example}

\section{Further analysis of the `trips abroad' data}
In this section we return to the analysis of the `trips abroad' data studied in \cite{NPES15}. Specifically, we look more closely at how the estimates for the log-spectrum differ depending on the data available to us. We restrict our attention to the cases in which we observe: 14 monthly counts of UK residents leaving the country; 28 quarterly counts; and the union of both these datasets. 

\begin{figure}
 \centering
\begin{subfigure}[b]{0.3\textwidth}
 \includegraphics[width=\textwidth]{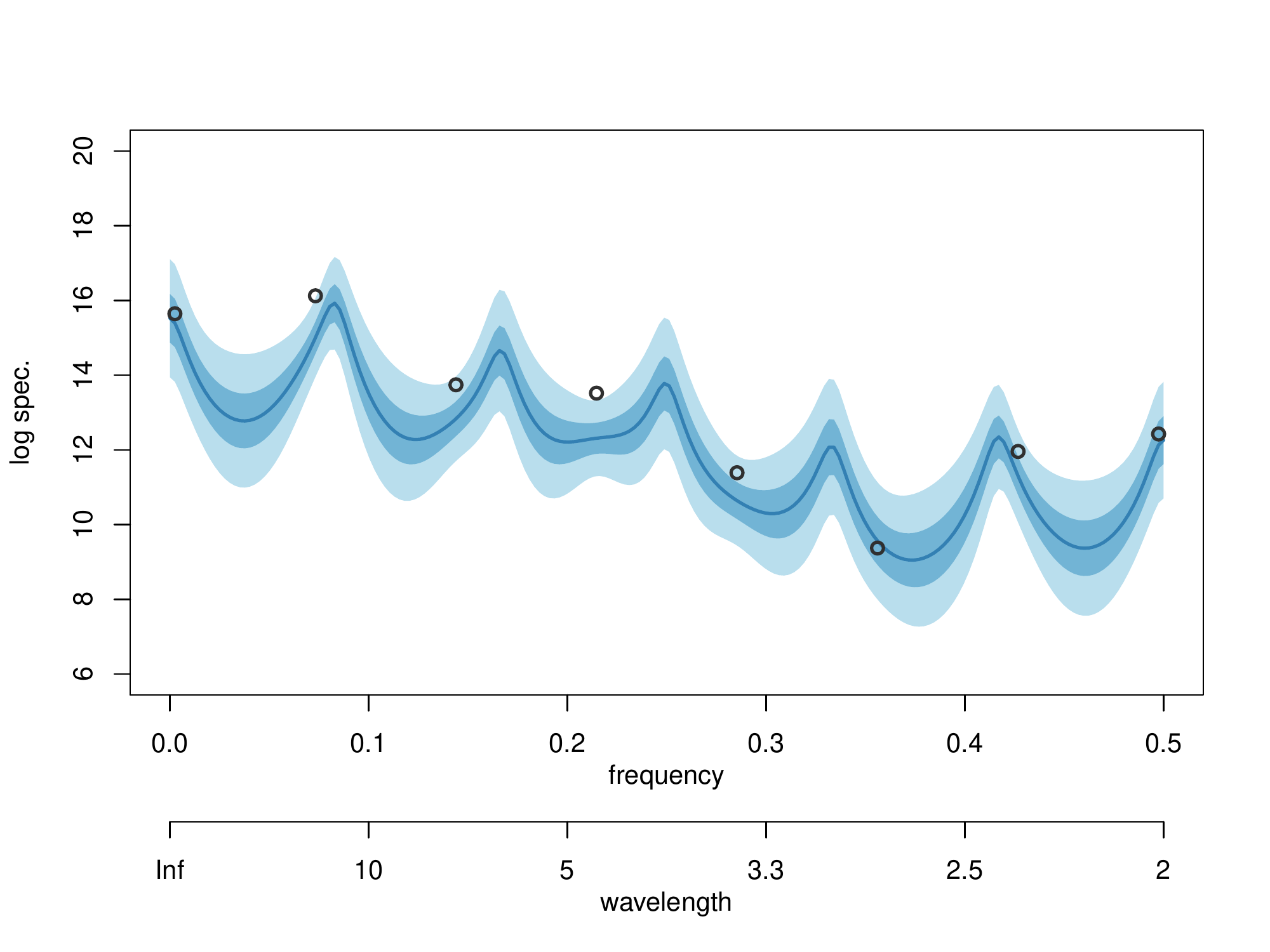}
\end{subfigure}
\begin{subfigure}[b]{0.3\textwidth}
 \includegraphics[width=\textwidth]{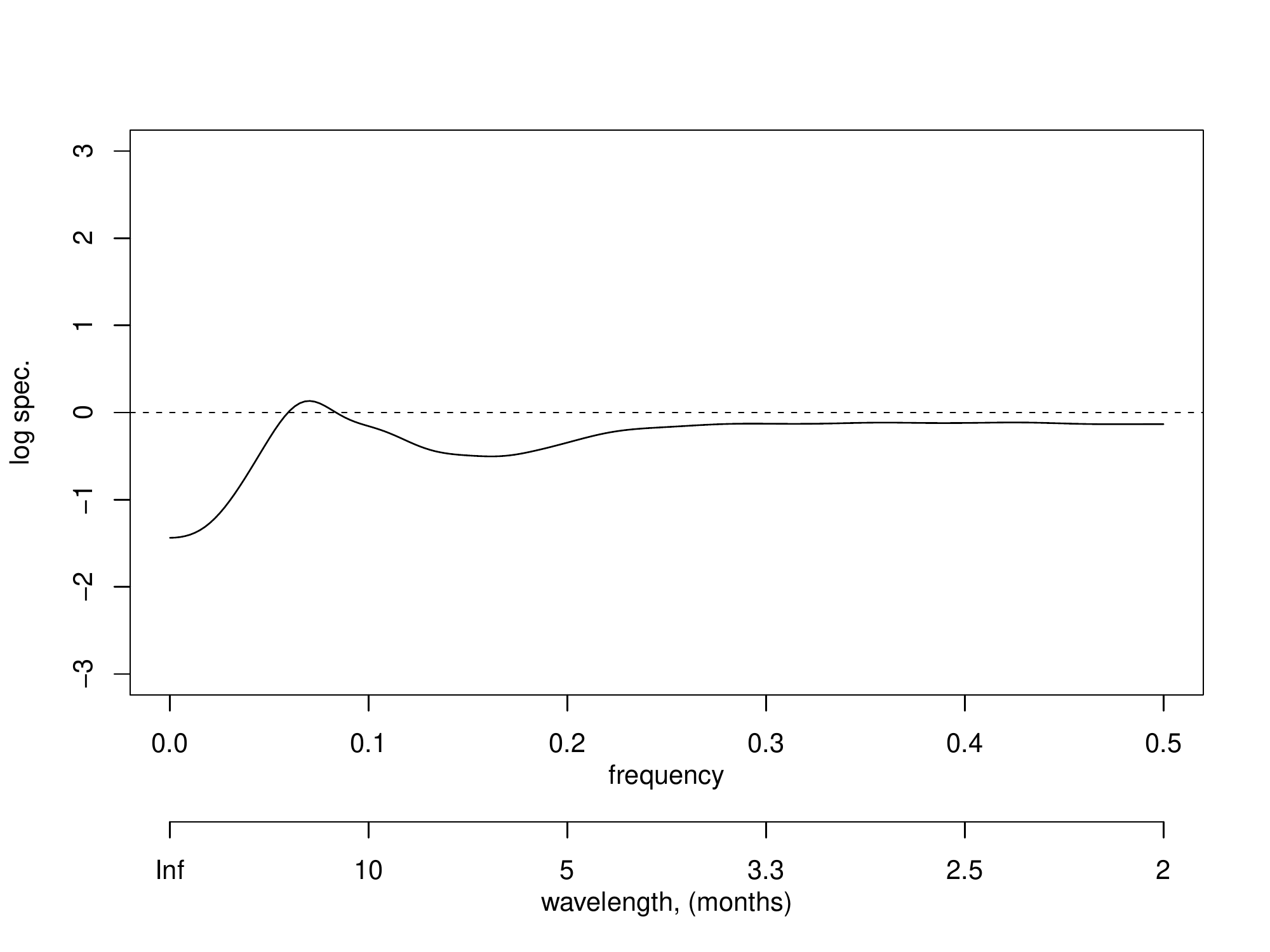}
\end{subfigure}
\begin{subfigure}[b]{0.3\textwidth}
 \includegraphics[width=\textwidth]{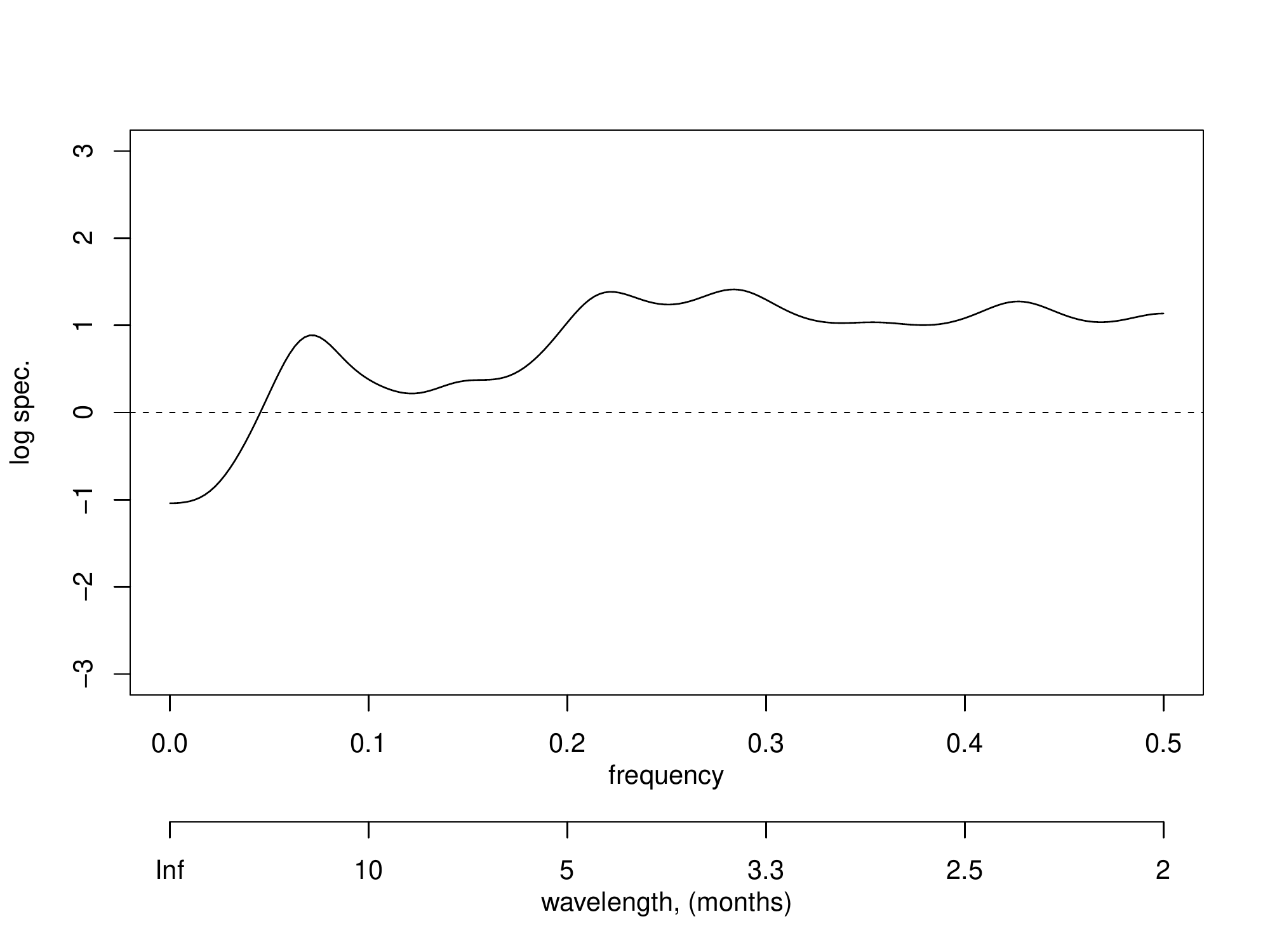}
\end{subfigure}\\
\begin{subfigure}[b]{0.3\textwidth}
\hfill
\end{subfigure}
\begin{subfigure}[b]{0.3\textwidth}
 \includegraphics[width=\textwidth]{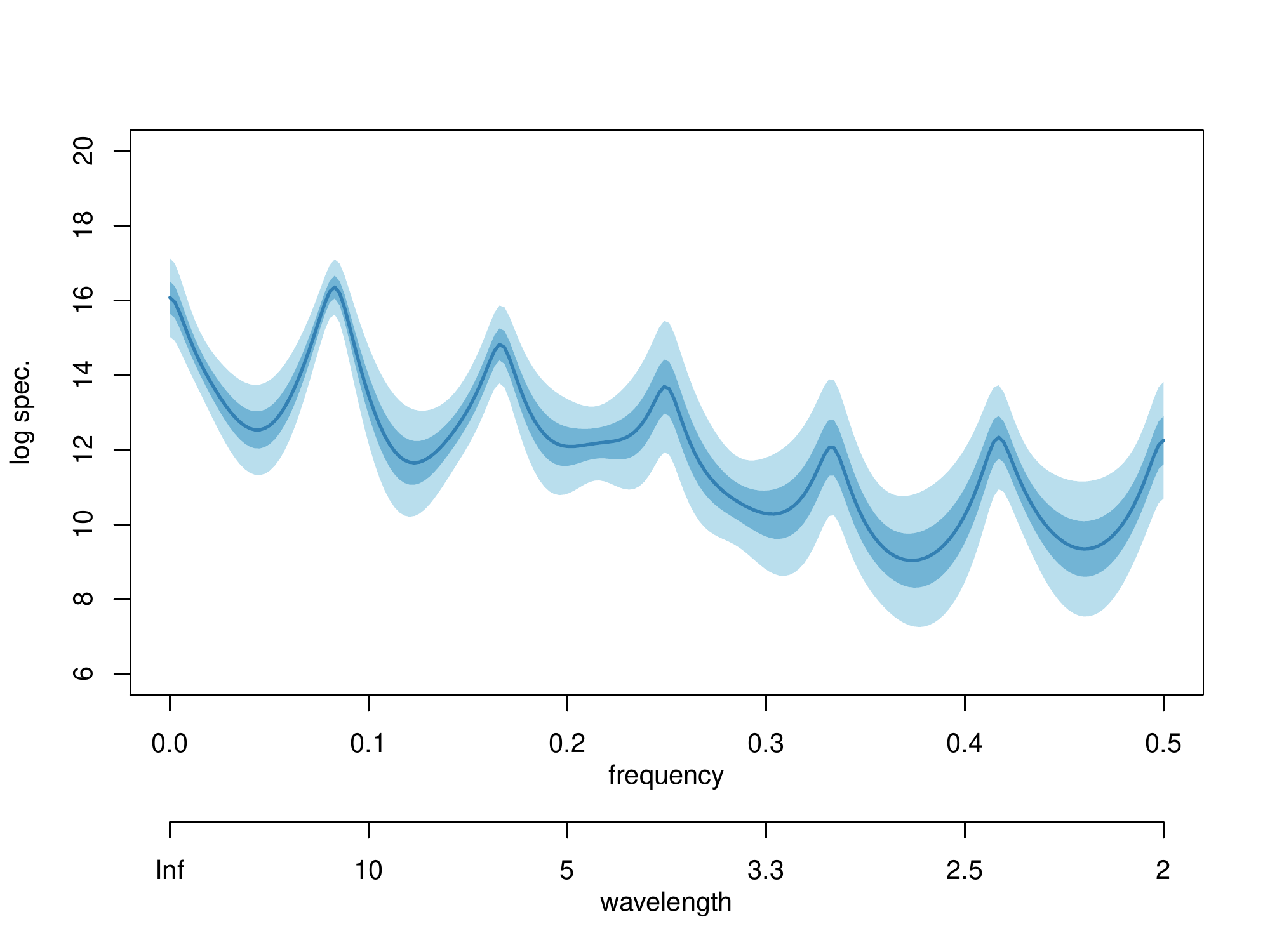}
\end{subfigure}
\begin{subfigure}[b]{0.3\textwidth}
 \includegraphics[width=\textwidth]{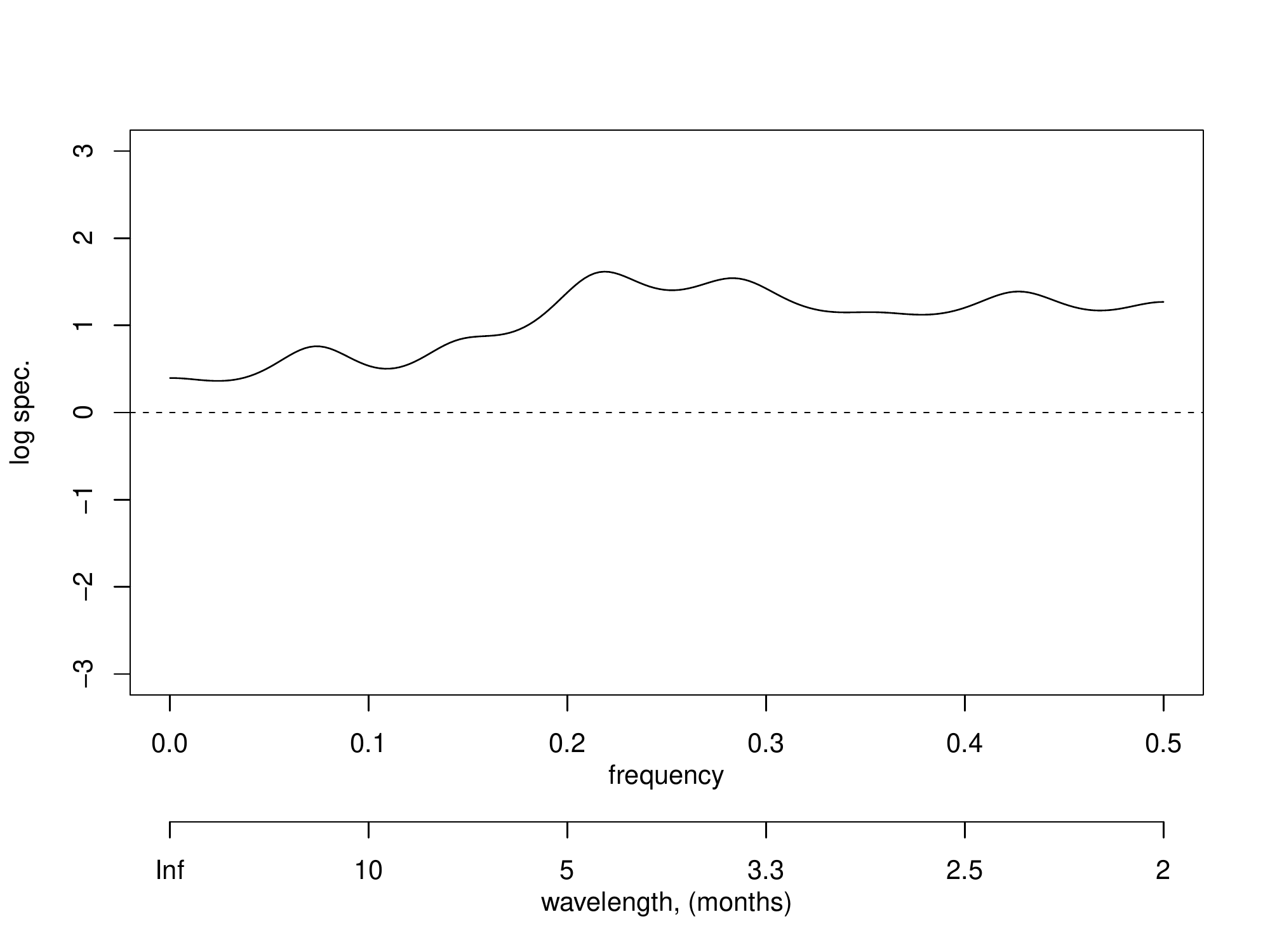}
\end{subfigure}\\
\begin{subfigure}[b]{0.3\textwidth}
\hfill
\end{subfigure}
\begin{subfigure}[b]{0.3\textwidth}
\hfill
\end{subfigure}
\begin{subfigure}[b]{0.3\textwidth}
 \includegraphics[width=\textwidth]{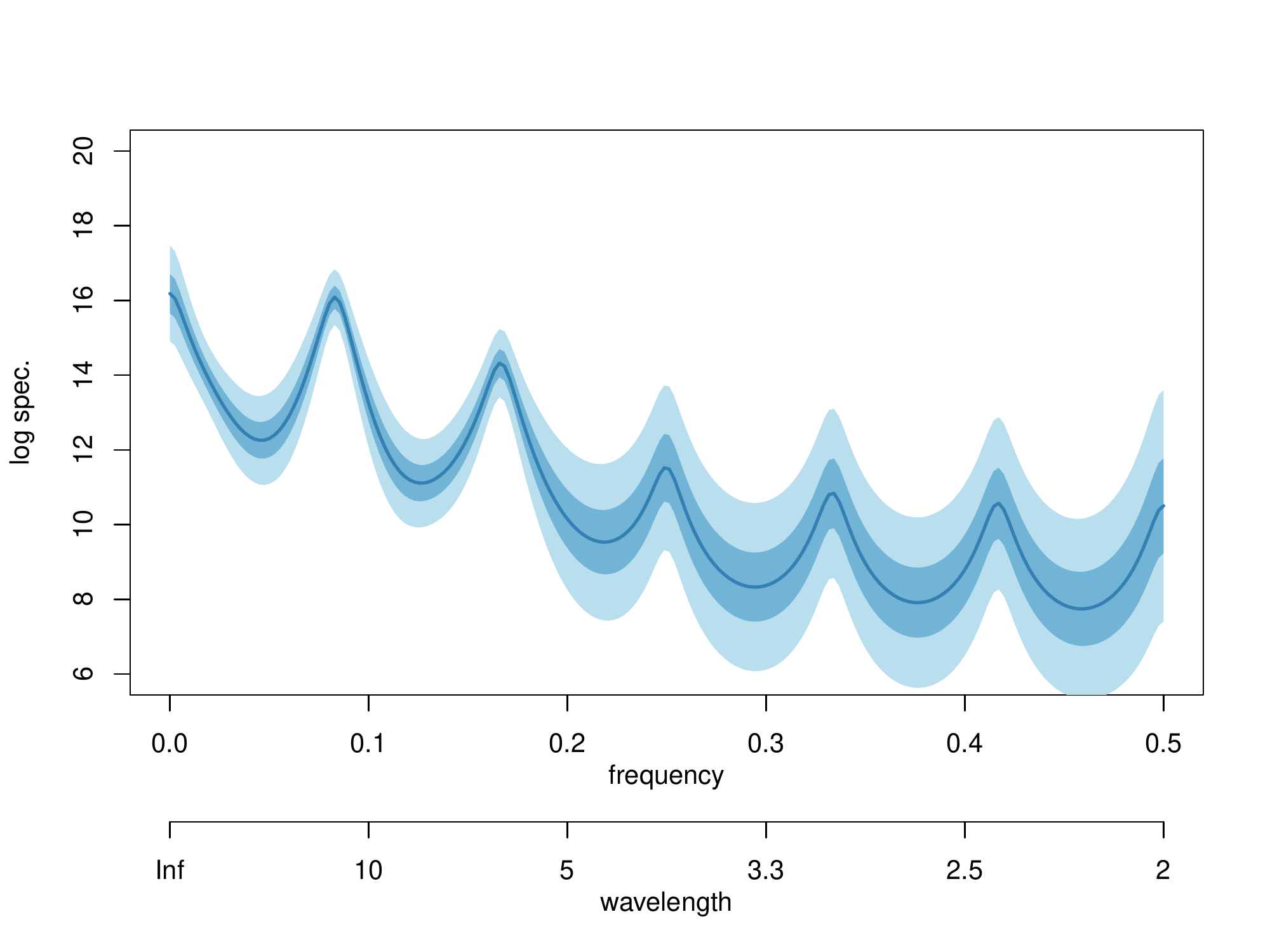}
\end{subfigure}
\caption{Log-spectra for the `trips aboard' data. In this array of plots the log-spectra based on the monthly data, the monthly and the quarterly data, and just the quarterly data are plotted on the diagonal in this order. The off-diagonals are the differences of means of the log-spectra. Note that the y-axes for the log-spectra and the differences of the log-spectra are different.}
\label{logspecdiffs}
\end{figure}

The estimates and the differences are plotted in Figure \ref{logspecdiffs}. We note, in particular, how the plotted differences show us that: learning about the quarterly data, having already seen the monthly data, causes us to adjust our estimate significantly at the lowest frequencies; and that learning about the monthly data, having already seen the quarterly data, causes us to adjust our estimate for all frequencies, but particularly the
higher ones above $0.1$.

\section{Communicating and propagating spectral uncertainty}
\begin{figure}
\centering
\makebox{\includegraphics[width=.8\textwidth]{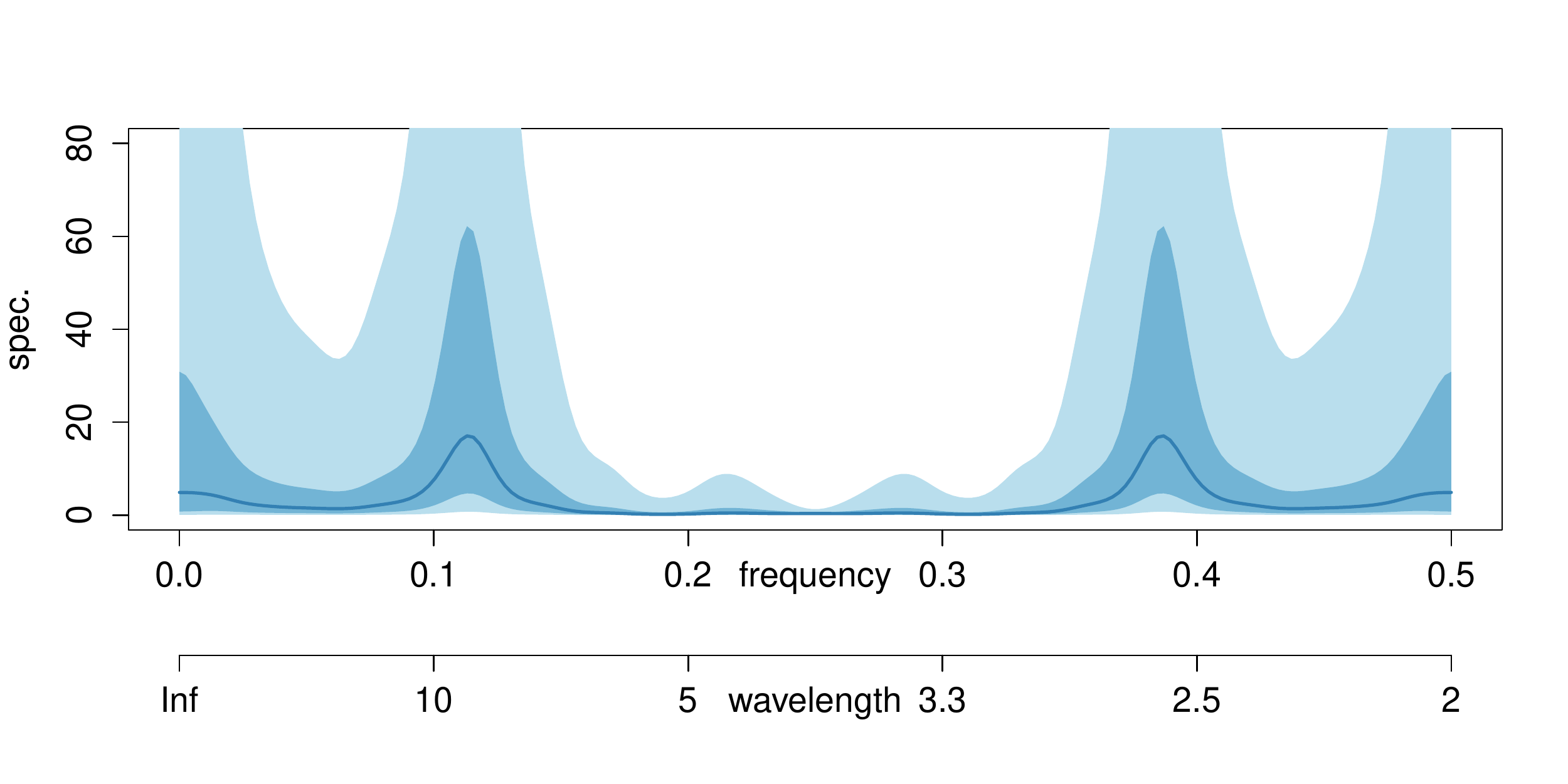}}\\
\makebox{\includegraphics[width=.8\textwidth]{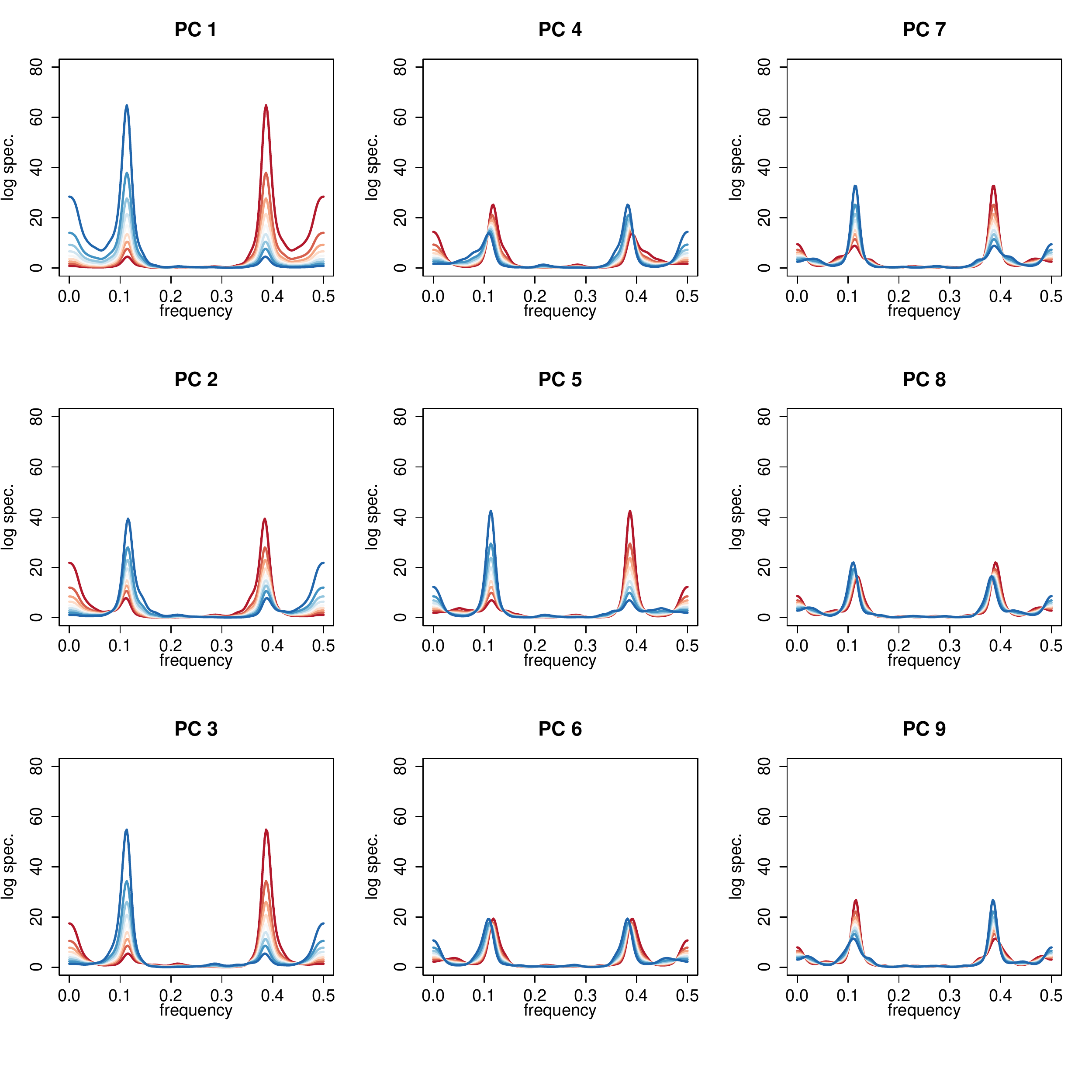}}
\caption{\label{pcs(every2)} In the top-most subplot we present exponentiated expectations and approximate $90\%$ and $50\%$ credible intervals for the log-spectrum of a process given only observations at every other time point. The aliasing phenomenon in this case is manifested in the symmetry around $0.25$. In the bottom grid of sublots we illustrate variation about the expectation in nine principal directions (in the log space) defined by the principal eigenvectors of the adjusted variance matrix for the log-spectrum basis coefficients.}
\end{figure}

\begin{figure}
\centering
\makebox{\includegraphics[width=.8\textwidth]{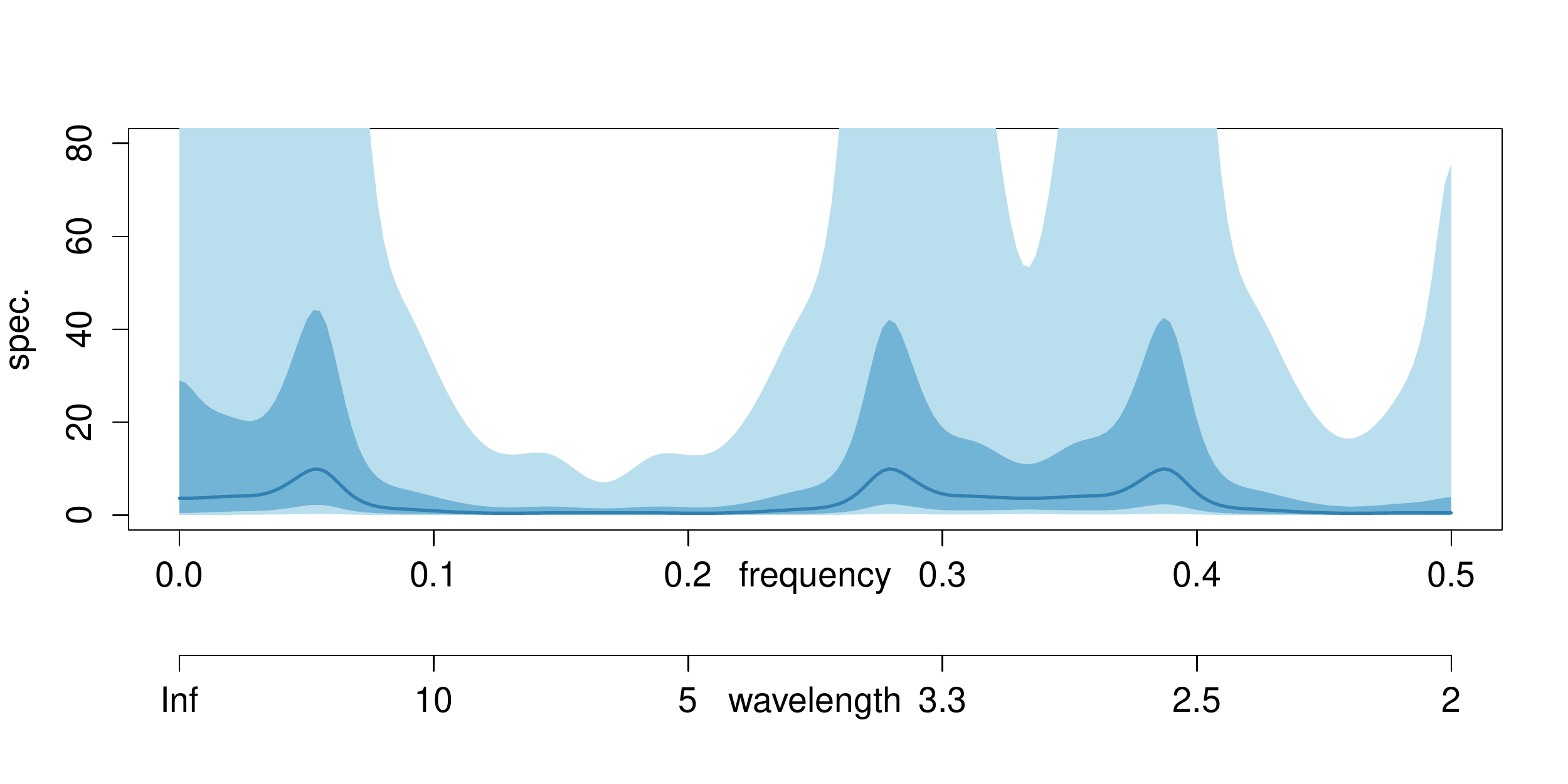}}\\
\makebox{\includegraphics[width=.8\textwidth]{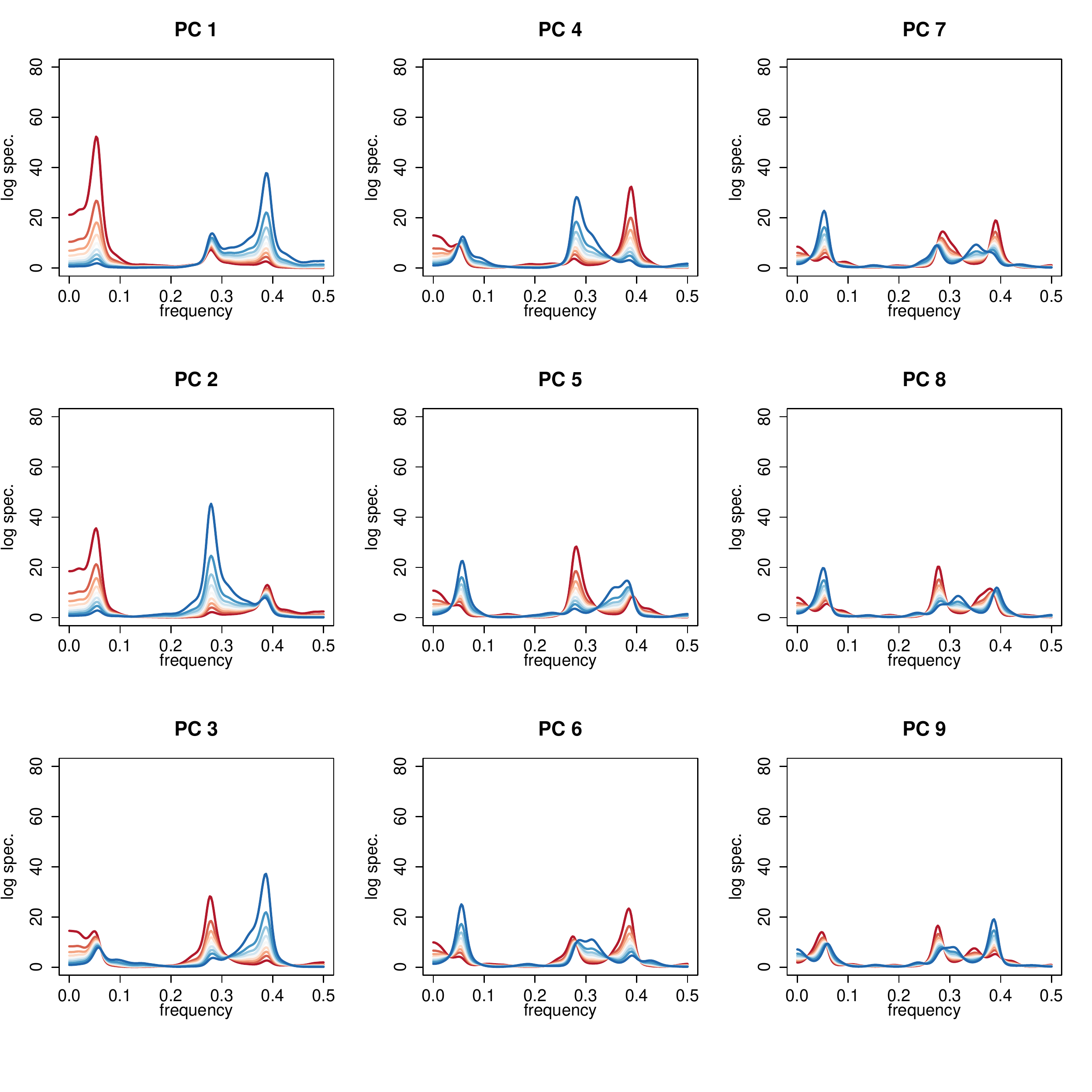}}
\caption{\label{pcs(every3)} These plots are entirely analogous to those in Figure \ref{pcs(every2)} except that the adjusted expectation and variance for the log-spectrum basis coefficients are informed by observations of the process at every third time point. Note that the symmetry of the moments is corrupted slightly by the smoothness penalty and the positioning of the reflecting/Nyquist frequencies.}
\end{figure}

The linear Bayes estimate for the log-spectrum of a stationary process, described in \cite{NPES15}, is accompanied by an adjusted variance matrix for the coefficients we use to parameterise it. This variance matrix leads trivially to a variance matrix for values of the log-spectrum. The diagonals of this matrix can then be used to derive approximate credible intervals for the log-spectrum at any particular frequency that can be plotted easily. The exponentiated bounds of approximate $90\%$ and $50\%$ credible intervals are plotted in upper-most subplots of Figures \ref{pcs(every2)} and \ref{pcs(every3)} for example. By stopping here, however, we neglect the off-diagonal elements that quantify covariances between log-spectrum values, covariances that encode statements along the lines of `if spectral power accumulates at point $x$, it does not accumulate at point $y$'. This sort of constrained uncertainty, often understood in terms of \textit{aliasing}, is particularly relevant when a significant portion of 
the data 
available 
to us consists of a historical series of subsampled process values.

To communicate more of the covariance information to problem stakeholders we can illustrate the dominant modes of uncertainty for the log-spectrum by drawing attention to projections of that uncertainty onto the first few principal components derived from the variance matrix of the log-spectrum's basis coefficients. In the lower grids of subplots in figures \ref{pcs(every2)} and \ref{pcs(every3)} we have produced plots doing this. To be precise, producing each element of the grid required that we calculate nine unit normal quantiles, which, after scaling by the relevant principal loading, we multiplied by copies of a particular principal component. To each of the resulting scaled copies of coefficient vectors we added the basis coefficients' expectation before multiplying them by matrices of basis function values. These constituted vectors of values of log-spectra that we then exponentiated and plotted. While these plots provide some understanding of the multidimensional nature of the spectrum uncertainty, 
they do 
not help us propagate the uncertainty forwards into subsequent calculations. To do this we suggest taking advantage of the sparse quadrature grid technology developed by \cite{BG04} and implemented in R by Jelmer Ypma. Specifically, we recommend using a sparse Gaussian quadrature grid to determine the first few principal component scores of an ensemble of log-spectra. Averaging over this ensemble, using the accompanying quadrature weights, serves as a way to approximately integrate out the spectrum uncertainty.

\section{Further discussion}
\subsection{A note on predictability}
An important implication of Kolmogorov's formula for the one step ahead prediction error of a stationary process,
\begin{align*}
 \var_{H}(X_{t})= \exp \left\{ 2 \int_{0}^{1/2} \log f(\omega) d\omega \right\},
\end{align*}
is that it is not the location of the spectral power that makes a process predictable. Rather the predictability depends on how spread out the power is. To appreciate this fact, it is helpful to consider two extreme cases: in the first, we have a maximally spread out, flat spectrum corresponding to a totally unforecastable white noise process; in the other case, we have finitely many atoms of spectral power corresponding to a process composed of finitely many sinusoids with random coefficients, which may, along with all future values of the process, be inferred precisely with finitely many observations. 

\subsection{Additional implications of the cost function}\label{costimplications}
An additional implication of cost function used in main decision problem of \cite{NPES15} is that predictions of the process in question are required only at every $K^{th}$ time point regardless of whether the sampling rate is increased. This means that process values chronologically between those that must be predicted, may be observed and learned from, but do not contribute to the loss.

Another assumption is that $N_{future}$, the length of a future forecasting period, is long enough for us to be able to infer the optimal forecast function quickly enough for the forecast losses attributable to using the wrong function to be negligible. Accordingly, in the following simulation experiment we calculated the average observed forecast errors using the true value of the spectrum. This means that when it came to evaluating the cost of a sampling strategy decision within the simulation, the expected forecast loss did not include additional losses associated with making predictions informed by the wrong autocovariance values.

Note also that while $N_{future}$ is assumed to be large, taking it to infinity would make the cost of choosing the sub-optimal sampling strategy infinitely more costly than the alternative, undermining any attempt to equate the value of information regarding this choice to the finite costs of trial data.

\end{spacing}
\end{document}